\documentclass[fleqn,usenatbib]{mnras}

\usepackage{newtxtext,newtxmath}
\usepackage[T1]{fontenc}
\usepackage{ae,aecompl}

\usepackage{graphicx}	% Including figure files
\usepackage{amsmath}	% Advanced maths commands
\usepackage{amssymb}	% Extra maths symbols
\usepackage{multirow}
\usepackage{ulem}
\title[Automatic Morphological Classification of Galaxies]{Optimising Automatic Morphological Classification of Galaxies with Machine Learning and Deep Learning using Dark Energy Survey Imaging}

\author[Ting-Yun Cheng et al.]{Ting-Yun Cheng,$^{1}$\thanks{E-mail: ting-yun.cheng@nottingham.ac.uk}
Christopher J. Conselice,$^{1}$ 
Alfonso Arag\'on-Salamanca,$^{1}$
Nan Li,$^{1}$ 
\newauthor
Asa F. L. Bluck,$^{2}$ 
Will G. Hartley,$^{3,4}$
James Annis,$^{5}$ 
David Brooks,$^{3}$ 
Peter Doel,$^{3}$ 
\newauthor
Juan Garc\'ia-Bellido,$^{6}$ 
David~J. James,$^{7}$ 
Kyler Kuehn,$^{8}$
Nikolay Kuropatkin,$^{5}$ 
\newauthor
Mathew Smith,$^{9}$ 
Flavia Sobreira,$^{10,11}$ 
and Gregory Tarle$^{12}$ 
\\
\\
$^{1}$School of Physics and Astronomy, The University of Nottingham, University Park, Nottingham, NG7 2RD, UK\\
$^{2}$Kavli Institute for Cosmology, The University of Cambridge, Madingley Road, Cambridge, CB3 0HA, UK\\
$^{3}$Department of Physics {\&} Astronomy, University College London, Gower Street, London, WC1E 6BT, UK \\
$^{4}$Department of Physics, ETH Zurich, Wolfgang-Pauli-Strasse 16, CH-8093 Zurich, Switzerland \\
$^{5}$Fermi National Accelerator Laboratory, P. O. Box 500, Batavia, IL 60510, USA\\
$^{6}$Instituto de Fisica Teorica UAM/CSIC, Universidad Autonoma de Madrid, 28049 Madrid, Spain\\
$^{7}$Harvard-Smithsonian Center for Astrophysics, Cambridge, MA 02138, USA\\
$^{8}$Australian Astronomical Optics, Macquarie University, North Ryde, NSW 2113, Australia\\
$^{9}$School of Physics and Astronomy, University of Southampton,  Southampton, SO17 1BJ, UK\\
$^{10}$Instituto de F\'isica Gleb Wataghin, Universidade Estadual de Campinas, 13083-859, Campinas, SP, Brazil\\
$^{11}$Laborat\'orio Interinstitucional de e-Astronomia - LIneA, Rua Gal. Jos\'e Cristino 77, Rio de Janeiro, RJ - 20921-400, Brazil\\
$^{12}$Department of Physics, University of Michigan, Ann Arbor, MI 48109, USA\\
}
\date{Accepted 2020 February 13. Received 2020 January 15; in original form 2019 March 02}

\pubyear{2020}

\begin{document}

\label{firstpage}
\pagerange{\pageref{firstpage}--\pageref{lastpage}}
\maketitle
\begin{abstract}
There are several supervised machine learning methods used for the application of automated morphological classification of galaxies; however, there has not yet been a clear comparison of these different methods using imaging data, or a investigation for maximising their effectiveness. We carry out a comparison between several common machine learning methods for galaxy classification (Convolutional Neural Network (CNN), K-nearest neighbour, Logistic Regression, Support Vector Machine, Random Forest, and Neural Networks) by using Dark Energy Survey (DES) data combined with visual classifications from the Galaxy Zoo 1 project (GZ1). Our goal is to determine the optimal machine learning methods when using imaging data for galaxy classification. We show that CNN is the most successful method of these ten methods in our study. Using a sample of $\sim$2,800 galaxies with visual classification from GZ1, we reach an accuracy of $\sim$0.99 for the morphological classification of Ellipticals and Spirals. The further investigation of the galaxies that have a different ML and visual classification but with high predicted probabilities in our CNN usually reveals an the incorrect classification provided by GZ1. We further find the galaxies having a low probability of being either spirals or ellipticals are visually Lenticulars (S0), demonstrating that supervised learning is able to rediscover that this class of galaxy is distinct from both Es and Spirals. We confirm that $\sim$2.5\% galaxies are misclassified by GZ1 in our study. After correcting these galaxies' labels, we improve our CNN performance to an average accuracy of over 0.99 (accuracy of 0.994 is our best result).
%This is a simple template for authors to write new MNRAS papers.
%The abstract should briefly describe the aims, methods, and main results of the paper.
%It should be a single paragraph not more than 250 words (200 words for Letters).
%No references should appear in the abstract.
\end{abstract}
%galaxies: general
%galaxies: spiral
%galaxies: elliptical and lenticular, cD
%galaxies: structure
%galaxies: morphologies
%methods: observational
%methods: statistical
%methods - data analysis.
%surveys

%%%%%%%%%%%%%%%%%%%%%%%%%%%%%%%%%%%%%%%%%%%%%%%%%%

%%%%%%%%%%%%%%%%% BODY OF PAPER %%%%%%%%%%%%%%%%%%
%This is a simple template for authors to write new MNRAS papers.
%See \texttt{mnras\_sample.tex} for a more complex example, and \texttt{mnras\_guide.tex}
%for a full user guide.

%All papers should start with an Introduction section, which sets the work
%in context, cites relevant earlier studies in the field by \citet{Others2013},
%and describes the problem the authors aim to solve \citep[e.g.][]{Lintott2011}.

%SECTION%
\section{Introduction}
The morphological classification of galaxies is a very important tool for understanding the history of galaxy assembly. It not only tells us about the evolution of galaxies, but it can also reveal the stellar properties of galaxies, and thus their histories. Since the pioneering work by \citet{Hubble1926}, nearby galaxies can be easily and clearly classified into two main types: early-type galaxies (ETGs), which include elliptical galaxies and lenticular galaxies, which are mostly massive, with older stellar populations, and no spiral structure; and late-type galaxies, which include spiral galaxies and irregular galaxies, often with spiral arms, and which consist of a younger population. These two types are the basic classifications of galaxies in local universe and have remained so for nearly a century.

Along with the data explosion by more and more survey projects in astronomy, e.g. The Sloan Digital Sky Survey (SDSS)\footnote{https://www.sdss.org}, the Large Synoptic Survey Telescope (LSST)\footnote{https://www.lsst.org}, the Dark Energy Survey (DES)\footnote{https://www.darkenergysurvey.org/} \citep{Abbott2018}, etc, which will image more than hundreds of millions of galaxies, the traditional manual classification analysis by experts is obviously impossible to deal with this enormous amount of data.

The series of the Galaxy Zoo projects \citep{Lintott2008, Lintott2011, Willett2013} are one of the most successful tool to solve the problem of large scale morphological analysis. It allows amateurs to do the classification by answering a series of questions based on galaxy images. However, classification analysis is complex and difficult such that background knowledge and experience are essential when doing it. In addition, while visual morphological classification with Galaxy Zoo is faster than for single individuals, it is also time-consuming. For example, the Galaxy Zoo Project spent around 3 years on obtaining the classifications of $\sim$300,000 galaxies, due to the need for so many individual classifications per object.  DES and LSST, for instance, would take on the order of $>100$ years to classify with the Galaxy Zoo project. Therefore, an efficient automated classification method by computational science is essential for the future of this field. The way forward is clearly through machine learning, although we are still learning the best ways to apply this to galaxy morphology and other areas of astronomy, e.g. star-galaxy separation \citep[][etc]{Odewahn1992, Weir1995, Ball2006}, the Galaxy Zoo challenge \citep{Chou2014}, the Strong Gravitational Lens Finding Challenge \citep{Metcalf2019}, etc. 

The concept and application of machine learning in computational science have been around for some time \citep{Fukushima1980}, and the application in astronomy started in the 1990s. However, it has not been widely used in astronomy until the last few years due to the big improvement of the computation ability of computers and the development of this technology. The first application of machine learning on morphological classification can be traced to \citet{Storrie-Lombardi1992}. They applied a neural network with an input layer of 13 parameters, e.g. stellar properties, brightness profile, etc., which gave an output of five different types of galaxies. Since then, a slew of studies in astronomy have appeared utilising the technology of machine learning \citep[e.g.][]{Huertas-Company2008, Huertas-Company2009, Huertas-Company2011, Shamir2009, Polsterer2012, Sreejith2018, Hocking2018}, neural networks \citep[e.g.][]{Maehoenen1995, Naim1995, Lahav1996, Goderya2002, Ball2004, Calleja2004, Banerji2010}, and Convolutional Neural Networks (CNN) \citep[e.g.][]{Dieleman2015, Huertas-Company2015, Huertas-Company2018, Dominguez-Sanchez2018} for the morphological classification of galaxies. 

There are now several different methods in machine learning used to carry out morphological classifications. However, although machine learning have been highly developed for decades there is not a clear quantitative comparison between these different methods yet especially concerning imaging data. In our study, we carry out a comparison of the simplest classification -- binary morphological classification of `Ellipticals' and `Spirals' (follows the classification of the Galaxy Zoo 1 project) -- between several common methods in machine learning (listed in Table~\ref{tab:methods}) using imaging data.

In previous studies, except for the application of CNN, there were very few studies which directly exploited imaging data when using other machine learning algorithms, such as neural networks or support vector machine. Therefore, we imitate the application of face and hand-writing recognition in computational science \citep{Bishop2006} that directly input image pixels as features to all the methods we compared for a fair comparison of different methods. 

In this study we use DES imaging data which has better resolution and deeper depth than SDSS images (see Section~\ref{sec:datasets}). With our machine learning algorithm, these properties of DES data help us to build a larger, deeper, and better catalogue of galaxy morphology containing the largest sample to date.  We therefore also discuss galaxies which 'fail' in our training algorithms, and discuss how these systems are often misclassified in Galaxy Zoo. We also discuss systems that have a low probability of being either an elliptical or a spiral and how these systems are visually classifiable on the DES imaging as lenticulars.  

The arrangement for this paper is as follows. Section~\ref{sec:datasets} describes the data resources, the procedure of pre-processing, and the datasets we use in this paper. The descriptions of each method are discussed in Section~\ref{sec:methods}. We present the main results in Section~\ref{sec:results} and include a further discussion in Section~\ref{sec:discussion}. The conclusion is shown in Section~\ref{sec:conclusions}.
%table
\begin{table}
	\centering
	\begin{tabular}{cl} % four columns, alignment for each
		\hline
		\multicolumn{1}{c}{Labels} & {Machine Learning Algorithms}\\
		\hline\hline
		\multicolumn{1}{c}{1} & {K-Nearest Neighbour (KNN)} \\
		\multicolumn{1}{c}{2} & {KNN + Restricted Boltzmann Machine} \\
		\multicolumn{1}{c}{} & {(KNN+RBM)} \\
		\hline
		\multicolumn{1}{c}{3} & {Support Vector Machine (SVM)} \\
		\multicolumn{1}{c}{4} & {SVM + Restricted Boltzmann Machine} \\
		\multicolumn{1}{c}{} & {(SVM+RBM)} \\
		\hline
		\multicolumn{1}{c}{5}  & {Logistic Regression (LR)} \\
		\multicolumn{1}{c}{6} & {LR + Restricted Boltzmann Machine} \\
		\multicolumn{1}{c}{} & {(LR+RBM)} \\
		\hline
		\multicolumn{1}{c}{7}  & {Random Forest (RF)} \\
		\multicolumn{1}{c}{8} & {RF + Restricted Boltzmann Machine} \\
		\multicolumn{1}{c}{} & {(RF+RBM)} \\
		\hline
		\multicolumn{1}{c}{9} & {Multi-Layer Perceptron Classifier} \\
		\multicolumn{1}{c}{} & {(MLPC)} \\
		\multicolumn{1}{c}{10} & {Convolutional Neural Network (CNN)} \\
		\hline
	\end{tabular}
	\caption{The list of machine learning methods tested in this study.}
	\label{tab:methods}
\end{table}
%SECTION%
\section{Data Sets}
\label{sec:datasets}
%This section introduces the resource of our data, the pre-processing detail, and the datasets for feeding to machine learning or deep learning algorithm. 
For the images in this analysis we use the subset of Dark Energy Survey (DES) Year 1 (Y1) GOLD data - DES observation of SDSS stripe 82, selected at magnitude \textit{i} $<$22.5 and redshift \textit{z} $<$0.7 \citep{Drlica-Wagner2018}. DES data covers 5000 square degrees ($\sim1/8$ sky) and partially overlaps with the survey area of the Sloan Digital Sky Surveys (SDSS), but has a better seeing than the SDSS images from Galaxy Zoo. Dark Energy Camera (DECam) \citep{Flaugher2015}, the new installed camera used in DES, which is mounted on the Victor M. Blanco 4-meter Telescope at the Cerro Tololo Inter-American Observatory (CTIO) in the Chilean Andes, improved the quantum efficiency in the infrared wavebands ($>$90\% from $\sim$650 nm to $\sim$900 nm), and gives a better quality images for the observation of very distant objects than previous surveys with the spatial resolution of ${ 0. }^{ \prime \prime  }263$ per pixel and the depth of $i=22.51$ \citep{Abbott2018}.

A DES survey image has more than 500M pixels. Each tile is 1/2 sq.-deg. The coadd (tile) images are 10000 by 10000 pixels in size with a pixel scale ${ 0. }^{ \prime \prime  }263$. The total number of the data in this subset is around 1.87 million galaxy stamps with photometric redshift, and photometry information in 308 \textit{i}-band coadd images.

In order to train our machine learning algorithm, we match the DES data with the visual morphological classifications from  the Galaxy Zoo 1 project (GZ1, hereafter)\footnote{https://data.galaxyzoo.org/} \citep{Lintott2008, Lintott2011}. we only exploit the visual classifications which have agreements (votes rates) over 80 percent and have been bias corrected by \citet{Bamford2009} for both Ellipticals and Spirals in GZ1. However, the matching of DES data with visual classifications from GZ1  only gives 2,862 objects in total, with the number ratio between Ellipticals and Spirals being 1 to 3. Their magnitude ranges from $\sim$12.5 to 18 in $i$-band, and the redshift z$\le$0.25 (peak at z$\sim$0.1). To avoid overfitting while carrying out the ML training, we apply data augmentation in the pre-processing procedure in our study (Section~\ref{sec:data_augmentation}). To improve the performance of our machine learning methods, we apply other techniques including feature extraction, i.e. Histogram of Oriented Gradient (HOG) \citep{Dalal2005} to extract other informative features from galaxy stamps (Section~\ref{sec:feature_extraction}).
%SUBSECTION%
\subsection{Pre-Processing}
\label{sec:pre_processing}
Before data pre-processing, we separate our 2,862 galaxies with DES data and the GZ1 classification randomly into training sets, and testing set, to prevent repeated galaxies in both sets. Our data pre-processing has four main steps: (1) Data Augmentation; (2) Stamps creation; (3) Feature Extraction; (4) Rescaling. The details are shown below.
\begin{figure*}{}
\begin{center}
\graphicspath{}
	%\hspace*{-4.5ex}
	%present the procedure of pre-processing
	\includegraphics[width=2.1\columnwidth]{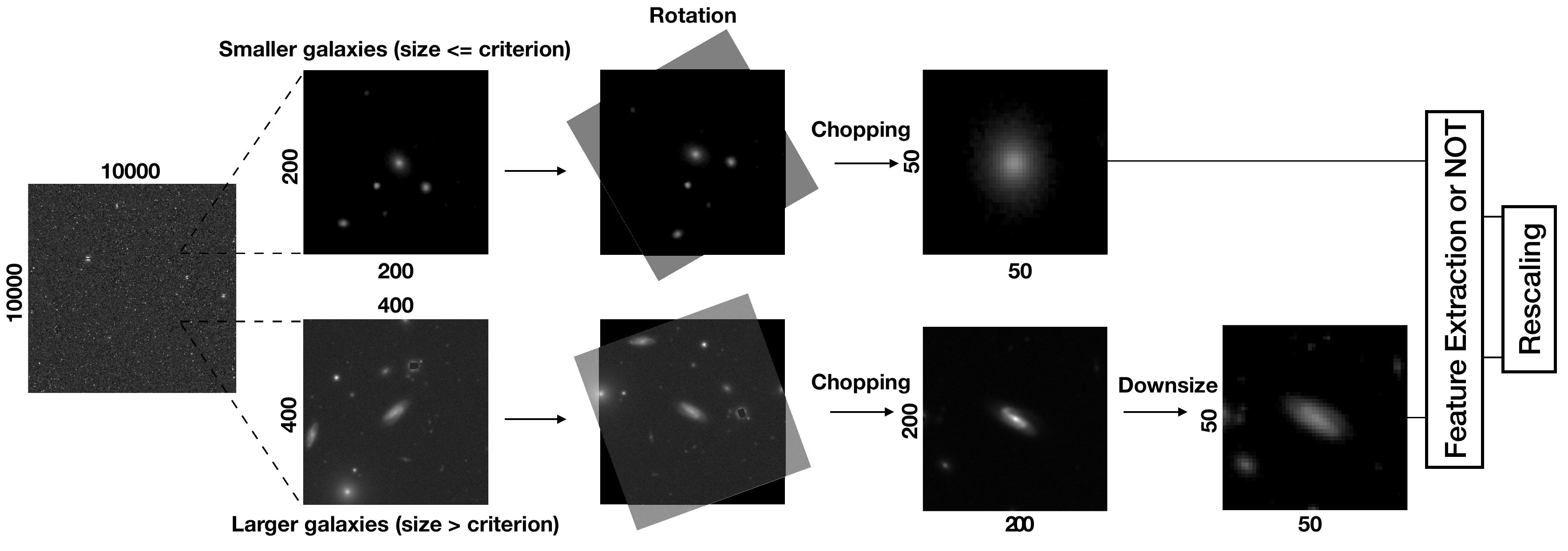}
   	\caption{Pre-processing procedure pipeline. The pipeline starts from the initial coadd images, then we chop the coadd images into different sizes according to the size of galaxies. After rotation, we chop and downsize the images to the required sizes: 50 by 50 pixels. The details of the procedure is in Section~\ref{sec:pre_processing}}
    	\label{fig:preprocessing}
\end{center}
\end{figure*}
%SUB-SUBSECTION%
\subsubsection{Data augmentation}
\label{sec:data_augmentation}
Data augmentation is of great importance while using pixel inputs in machine learning. Since \citet{Dieleman2015}, data augmentation by rotating images has been widely used within CNN for the morphological classification of galaxies. In this paper, we have 2,862 galaxies with visual classifications from GZ1, 759 Ellipticals and 2,103 Spirals, respectively, to train and test our methods. In order to prevent over-fitting during training, we rotate each galaxy image by 10 degrees differences from 0 to 350 degrees to increase the number of training samples. Hence, the available number of training samples increase to $\sim$100,000. After rotation, we add Gaussian noise to the rotated images \citep{Huertas-Company2015}.  This noise is small enough to not to influence the visual appearance and structures of the galaxies (namely, remain the same visual classification), but it is big enough to make a detectable but change of pixel values.

Although data augmentation through rotating images is a well known method used in machine learning application \citep[e.g.][]{Dieleman2015, Huertas-Company2015}, the effect of these rotated images is unexplored. Therefore, we investigate the difference of performance between partially and fully using rotated images in the datasets in Section~\ref{sec:rotated_images}.

%SUB-SUBSECTION%
\subsubsection{Creation of the galaxy stamps}
\label{sec:stamp_creation}
Fig.~\ref{fig:preprocessing} shows the pre-processing procedure used in our study. Using the galaxy catalogue from DES, we cut the coadd images with units of size 10000 by 10000 pixels into millions of galaxy stamps with sizes of 50 by 50 pixels. The size of galaxy stamp is based on the size distribution of galaxies in the DES Y1 GOLD data (stripe 82), where over 99{\%} of galaxies are smaller than a threshold of 25 by 25 pixels. Therefore, the size of our stamp is 50 by 50 pixels, which is twice as large as the threshold in the size distribution of galaxies.

Fig.~\ref{fig:preprocessing} shows that before chopping the stamp to the size of 50 by 50 pixels, we create the galaxy stamps with an initial size of 200 by 200 pixels when the galaxy size is smaller than 30 by 30 pixels, and 400 by 400 pixels when the galaxy size is larger than 30 by 30 pixels. For smaller galaxies, we rotate the 200 by 200 pixels stamps first, then reduce them in size to 50 by 50 pixels; for larger galaxies, we rotate 400 by 400 pixel stamps, reduce them in size to 200 by 200 pixels, then downsize them to 50 by 50 pixels by calculating the mean value of pixels in a size of 4 by 4 pixel cell. This procedure is designed to prevent empty pixel values showing up at the corner of stamps when we rotate images with non-90 degrees rotations.

%SUB-SUBSECTION%
\subsubsection{Feature Extraction}
\label{sec:feature_extraction}
\begin{figure}
\begin{center}
\graphicspath{}
	%HOG example
	\includegraphics[width=0.8\columnwidth]{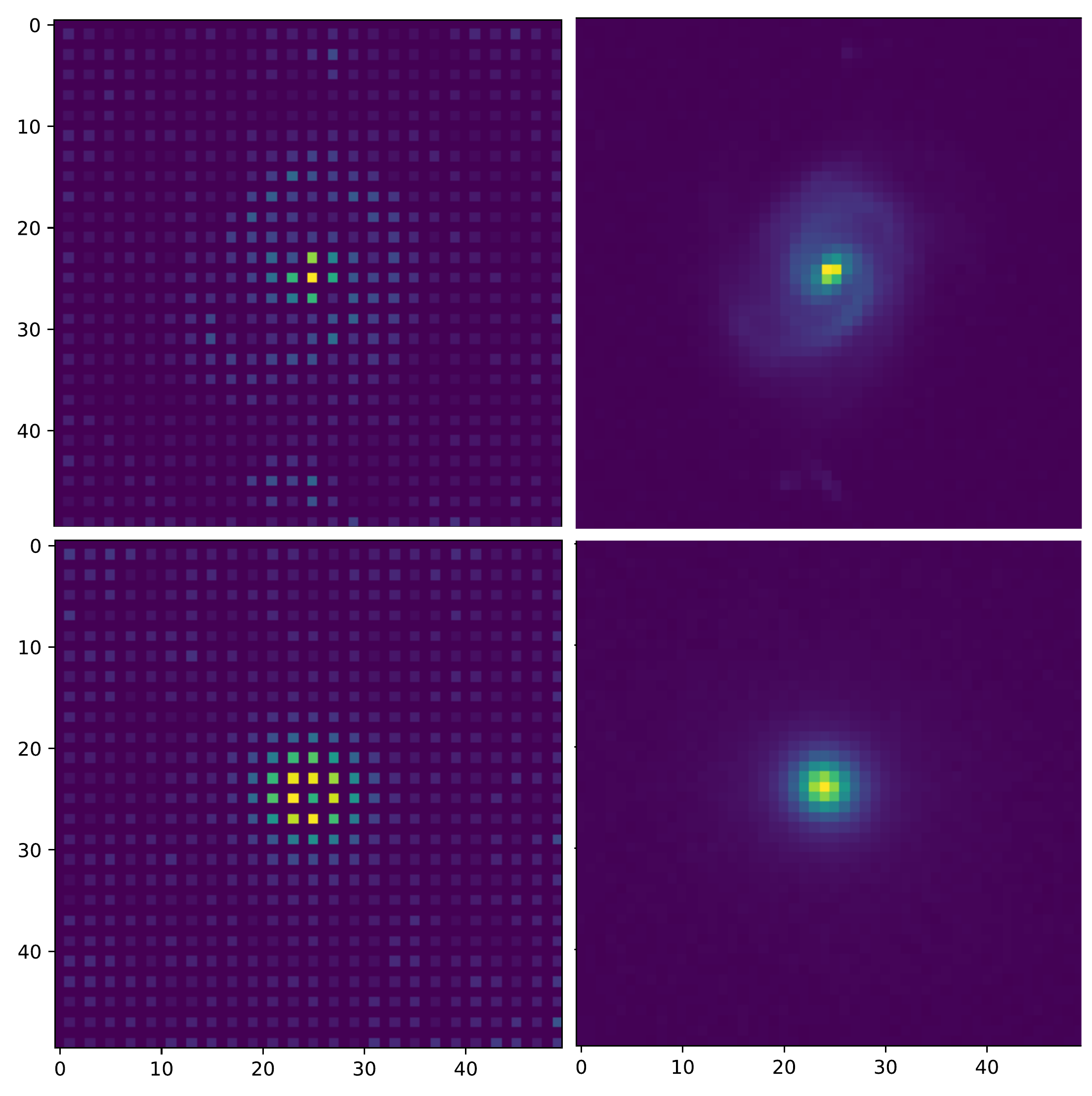}
   	\caption{Examples of images from Histogram Oriented Gradient (HOG) with the cell size of 2 by 2 pixels. {\it{Left}}: HOG images. {\it{Right}}: original images in linear scale. {\it{Top}}: Spirals. {\it{Bottom}}: Ellipticals.}
    	\label{fig:hog_example}
\end{center}
\end{figure}

In our study, we apply the Histogram of Oriented Gradients (HOG) on both our original and rotated stamps to investigate the impact of this feature extractor on supervised machine learning. HOG is a feature extractor which is able to extract the distribution of gradients with their direction from each pixel value. It is useful for characterising the appearance and the shape of objects \citep{Dalal2005}. It calculates the gradients of the horizontal (x) and vertical (y) direction of stamps. The magnitude and orientation of the gradient are calculated as below,
%equation
\begin{equation}
    	\left| G \right| =\sqrt { { G }_{ x }^{ 2 }+{ G }_{ y }^{ 2 } }, \\ \theta =\arctan { \left( \frac { { G }_{ y } }{ { G }_{ x } }  \right)  } 
	\label{eq:hog}
\end{equation}
%paragraph
where $\left| G \right|$ is the gradient magnitude of each pixel, ${ G }_{ x }$ is the gradient magnitude measured in x-direction, ${ G }_{ y }$ is the gradient magnitude measured in y-direction, and $\theta$ is the orientation of the gradient for each pixel in the images. It then measures the contribution of gradients from each pixel in the cell with the size of 2 by 2 pixels, and uses a histogram to describe the contribution of gradient magnitude to each orientation of gradient. The input of HOG image is the direct output of this feature extraction process, and we rescale the pixel value to the range between 0 and 1 (Section~\ref{sec:rescale}). Examples of HOG images are shown in Fig.~\ref{fig:hog_example}.

HOG is very popular within pattern recognition studies, e.g. human detection, face recognition, and handwriting recognition \citep[e.g.][etc]{Dalal2005, Shu2011, Kamble2015}; however, it is not popular yet in astronomy studies for the usage of machine learning algorithms. One of the applications is the detection of gravitational lensing images \citep[]{Avestruz2019}, and a few previous works on the galaxy morphology \citep[e.g. The Galaxy Zoo challenge ][]{Chou2014}. However, none of these studies have examined the influence of HOG on the performance of machine learning algorithms. In this study, we apply HOG on our images to investigate not only the effect of it on automated morphological classification of galaxies, but also the impact of it on the performance of different machine learning algorithms (Section~\ref{sec:different_input}).
%figure

%SUB-SUBSECTION%
\subsubsection{Rescaling}
\label{sec:rescale}
Rescaling is a very important process in the application of machine learning. Different galaxies have different brightness due to their different properties and their distances, so the pixel values of each image have significant variation between galaxies. This would cause difficulties for machine learning algorithms when defining the boundaries between different classes. Therefore, we rescale the pixel values of each image (raw and HOG images) to the range between 0 and 1 through normalising by the maximum and minimum pixel value of each image. We are aware that intrinsic brightness can be a classification criteria, including surface brightness.  However, in this study we are interested in the structure only and not on other properties that might correlate with a class of galaxy such as surface brightness.
%SUBSECTION%
\subsection{The datasets}
\label{sec:the_datasets}
In this study, we create 4 different datasets (see Table~\ref{tab:datasets}). The first two datasets (1 $\&$ 2) contain both the original images and the rotated images, and the last two (3 $\&$ 4) contain only the rotated images. This setting is used for investigating the influence of rotated images on the performance (Section~\ref{sec:rotated_images}).

On the other hand, the datasets 1 $\&$ 3 are unbalanced which contain more spiral galaxies than elliptical galaxies in the datasets while the datasets 2 $\&$ 4 have an equal number of spiral galaxies and elliptical galaxies in each dataset. We balance the number of each type by adding different numbers of rotated images to each type. For example, we rotate images of the Ellipticals 7 times, but only 2 times for the images of Spirals in dataset 2, and 3 times for both types in dataset 1. We use this setting to investigate the effect of the balance between the number of each type in training samples (Section~\ref{sec:balance}). In addition, we also reduce the differences in the number of total training samples between each dataset to reduce the probable bias from this.

On the other hand, we have 2 (or 3 in CNN) different types of input data (i,  ii,  iii). The first type (i) is the raw image with linear scale, and the second type (ii) is the HOG image from feature extraction. The third type, `combination input (iii)', is special for CNN due to the characteristic structure of CNN that we can combine both the raw images (i) and HOG images (ii) as input without increasing the number of features. This is an new way to combine data using CNN whereas people used to restore the images with different colours in the third dimension of CNN in previous studies. We then also investigate the effect of this combination input (iii) and compare it with the other two types (i $\&$ ii) (Section~\ref{sec:different_input}).

For the testing set, we randomly pick 500 galaxies from 2,862 galaxies for each type (Ellipticals and Spirals). The rest of unselected galaxies are training set. Therefore, we have 1,000 galaxies in total for testing and the ratio between Ellipticals and Spiral is 1:1.
%table
\begin{table}
	\centering
	\begin{tabular}{cccc} % four columns, alignment for each
		\hline
		labels & i (raw), & ii (HOG), & iii (combination, for CNN)\\
		\hline\hline
		\multicolumn{1}{c|}{1} &\multicolumn{2}{l}{original images+rotated images} &\multicolumn{1}{l}{E:S$\sim$1:3, Training$=$10,448}\\
		\multicolumn{1}{c|}{2} &\multicolumn{2}{l}{original images+rotated images} &\multicolumn{1}{l}{E:S$\sim$1:1, Training$=$11,381}\\
		\multicolumn{1}{c|}{3} &\multicolumn{2}{l}{only rotated images}                          &\multicolumn{1}{l}{E:S$\sim$1:3, Training$=$11,448}\\
		\multicolumn{1}{c|}{4} &\multicolumn{2}{l}{only rotated images}                          &\multicolumn{1}{l}{E:S$\sim$1:1, Training$=$12,381}\\
		\hline
	\end{tabular}
	\caption{The arrangement of training datasets in this paper. The content included in the datasets are shown in the second column, and the third column shows that the ratio between Ellipticals and Spirals and the total number of training data in each dataset.}
	\label{tab:datasets}
\end{table}
%SECTION%

\section{Models of Machine Learning}
\label{sec:methods}
The concept of machine learning can connect with the invention of calculators \citep{Turing1950} that we program machine to obtain the information we want through the input numbers or characters (features). The breakthrough of visual pattern recognition in machine learning started from \citet{Fukushima1980} which proposed a hierarchical and multilayered neural network - Neocognitron. Machine learning stood on the stage of astronomical applications since the 1990s \citep[e.g.][etc]{Odewahn1992, Storrie-Lombardi1992, Weir1995}. 

There are two main types of features, `parameter input' and `pixel input', that can be fed into machine. In the studies of galaxy morphological classification, the `parameter input' is where we use parameters, which have clear correlations with galaxy types \citep[e.g.][]{Storrie-Lombardi1992, Naim1995, Lahav1996, Ball2004, Huertas-Company2008, Huertas-Company2009, Banerji2010, Huertas-Company2011, Sreejith2018}. For example, the `parameter' input can be surface brightness profile, colour, C-A-S system \citep{Conselice2003}, Gini Coefficient \citep{Abraham2003}, etc. 

On the other hand, the `pixel input' means that we treat each pixel of an image as a feature to feed machine learning algorithms. The `pixel input' is the most straightforward feature used in two for machine to learn although it significantly increases the number of features for computation. However, it is uncommon in previous studies of automated classification of galaxy morphology to use `pixel input' \citep[e.g.][]{Maehoenen1995, Goderya2002, Calleja2004, Polsterer2012} until the application of CNN become popular in recent years \citep{Dieleman2015, Huertas-Company2015, Dominguez-Sanchez2018}.

We use `pixel input' for each method in this study to investigate the effect of `pixel input' on different machine learning algorithms (Table~\ref{tab:methods}). Restricted Boltzmann Machine (RBM) \citep[][]{Smolensky1986, Hinton2002, Salakhutdinov2007}, shown in Table~\ref{tab:methods}, is the simplest neural network with one hidden layer, which we treat as a feature extractor for in this study (Section~\ref{sec:RBM}).

All of the codes in this study are built on {\bf{Python}}. The main packages we use in this paper are {\bf{scikit-learn}\footnote{http://scikit-learn.org/stable/}} \citep{Pedregosa2012} for most of methods; {\bf{Theano}\footnote{http://deeplearning.net/software/theano/}}, {\bf{Lasagne}\footnote{http://lasagne.readthedocs.io/en/latest/}}, and {\bf{nolearn}\footnote{https://pythonhosted.org/nolearn/}} for CNN.

%SUBSECTION%
\subsection{Restricted Boltzmann Machine (RBM)}
\label{sec:RBM}
Restricted Boltzmann Machine (RBM) \citep[][]{Smolensky1986, Hinton2002, Salakhutdinov2007} contains one hidden layer which is the simplest neural network architecture (more explanation for the architecutre of neural network in section~\ref{sec:MLPC}). This is a useful algorithm for dimensionality reduction and feature learning; therefore, in this paper, the RBM is used as a feature extractor to connect each feature. It extracts the features which are more interlinked with each other before we feed them to other machine learning algorithms. The combination of machine learning algorithms such as logistic regression \citep{Chopra2017} and RBM is actually widely used in face and handwriting recognition. 

In this study, the setting of RBM is identical amongst all methods that we apply a fixed learning rate ($=$0.001), 1,024 numbers of hidden units, and 500 iterations for RBM in training, where the learning rate determines how far to move the weights each time towards the local minimum of loss function. The number of iteration is approximately determined by where the maximum of log-likelihood is shown.

%SUBSECTION%
\subsection{k-Nearest Neighbours (KNN)}
\label{sec:KNN}
K-Nearest Neighbours (KNN) is the simplest non-parametric machine learning algorithm \citep[][]{Fix1951, Cover1967, Short1981, Cunningham2007}. This is one of the most common methods in pattern recognition and has several applications in clustering and classification problems \citep[in astronomy e.g.][]{Kugler2015}. The concept of KNN is to find highly similar data, where similarity is defined by the `distance' in the feature space between data. Parameter {\it{k}} is the number of nearest neighbours counted in the same group. This factor controls the shape of the decision boundary for the distribution of data. 

Increasing the value of {\it{k}} decreases the variance in the classification but also increases the bias of the classification. We chose the value of {\it{k}} by plotting the accuracy (Equation~\ref{eq:accuracy}) versus different values of {\it{k}}, and the value we ultimately use is {\it{k}=}5. The distance metric for calculating the distance between each data is defined by the {\it{Euclidean metric}}.%{\it{Minkowski metric}},

%SUBSECTION%
\subsection{Logistic Regression (LR)}
\label{sec:LR}
Logistic Regression (LR) is a generalised linear model \citep[][]{McCullagh1989} which uses the sigmoid function $\frac { 1 }{ 1+{ e }^{ -x } } $ (or logistic function) to output the probability of classification. The application in astronomy such as \citet{Huppenkothen2017} studies the variability of galactic black hole binary. 

The combination of LR and RBM is commonly used in face and handwriting recognition \citep[][]{Chopra2017}. The improvement of this combination is rather significant in LR while using `pixel input' because of the characteristics of neural networks (See section~\ref{sec:results}). 

%SUBSECTION%
\subsection{Support Vector Machine (SVM)}
\label{sec:SVM}
The concept of Support Vector Machine (SVM) algorithm is to find a hyperplane with the maximal distance to the nearest data for each type ({\it{support vector}}) \citep{Vapnik1995, Cortes1995}. In this study, we use a non-linear SVM, in particular, the Radial Basis Function (RBF) kernel function \citep{Orr1996}: $\left( { \vec{x} },\vec{{ x }}^{ ' } \right) \rightarrow K\left( { \vec{x} },\vec{{ x }}^{ ' } \right) =exp\left( -\gamma { \left\| { \vec{x} }-{ \vec{x} }^{ ' } \right\|  }^{ 2 } \right)$. The detailed introduction of the SVM algorithm is given in Appendix~\ref{sec:SVM_appendix}.

SVM was expecting to be an alternative option for the neural network due to the capability of dealing with high-dimensional data \citep{Zanaty2012}. The application of this in astronomy is very popular, e.g. \citet{Gao2008, Huertas-Company2008, Huertas-Company2009, Kovacs2015}. In this study, we use Nu-SVM which was first introduced by \citet{Scholkopf2001}, and apply the {\bf{Python}} package {\bf{NuSVC}}. The value of nu is determined by the {\bf{Python}} package {\bf{GridSearchCV}} \citep{Hsu2003}.

%SUBSECTION%
\subsection{Random Forest (RF)}
\label{sec:RF}
Random Forest (RF) is an ensemble learning method developed by \citet{Breiman2001} which aggregates the results from a number of individual decision trees to decide the final classification \citep{Fawagreh2014}. Each tree is trained by a randomly picked subset from the training set. The RF is a well known machine learning technique applied in Astronomy using `parameter input' \citep[e.g.][]{Dubath2011, Beck2018} but the application that directly using pixel such as our study is untested. 

We use {\bf RandomForestClassifier} from the {\bf scikit-learn} module \citep{Pedregosa2012}. The number of trees ({\bf n{\_}estimators}) used in this study is determined by plotting the accuracy (Equation~\ref{eq:accuracy}) versus different values of {\bf n{\_}estimators}, and we ultimately use 200 trees. The maximal number of features to consider for each split ({\bf max{\_}features}) is equal to $\sqrt { { N }_{ f } }$, where ${{ N }_{ f }}$ is the total number of features. Each tree grows until all leaves are pure or all leaves contain the number of leaves less than 2.

%SUBSECTION%
\subsection{Multi-Layer Perceptron Classifier (MLPC)}
\label{sec:MLPC}
\begin{figure}
\begin{center}
\graphicspath{}
	%MLPC structure introduction
	\includegraphics[width=0.6\columnwidth]{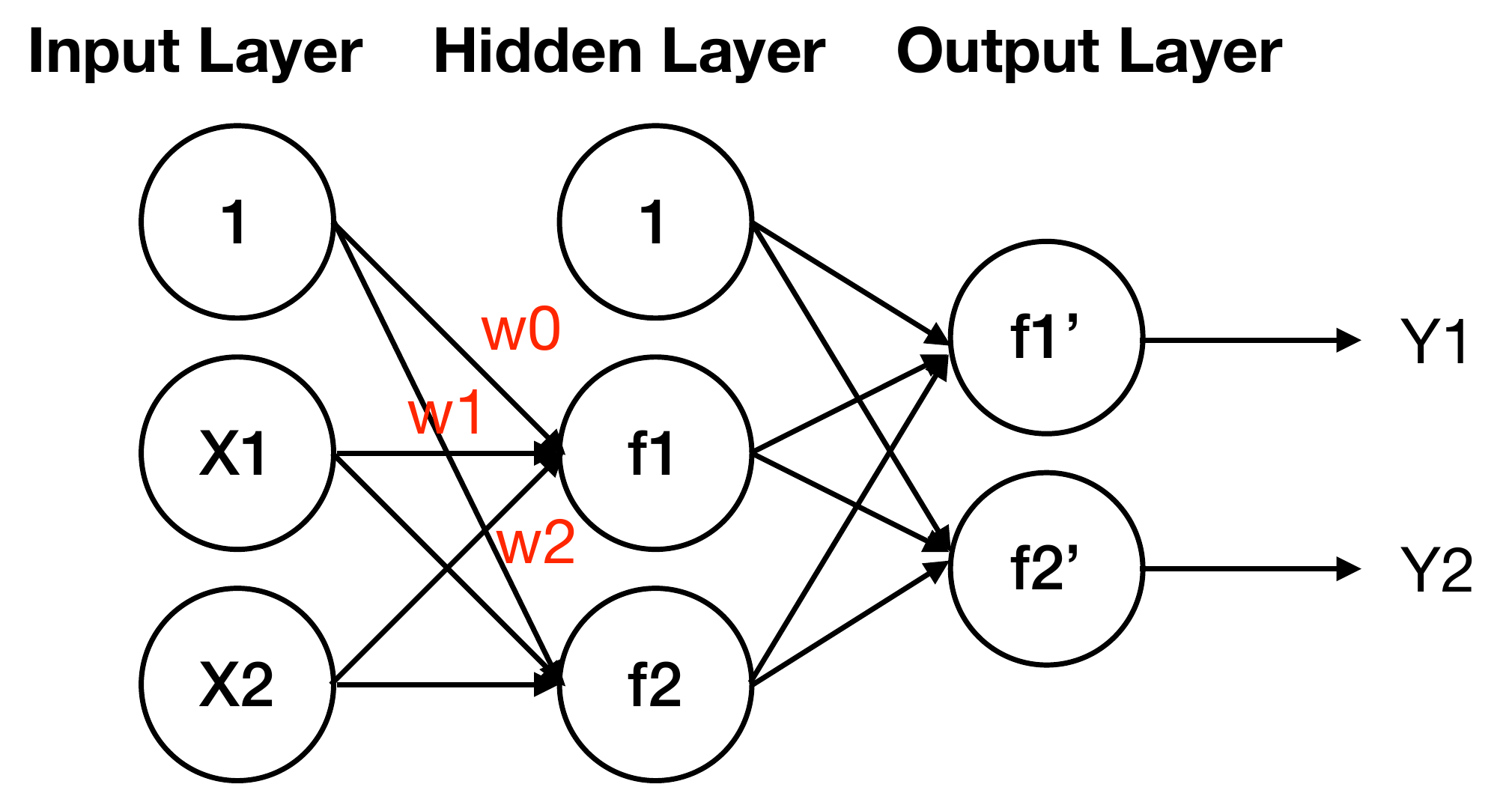}
   	\caption{Illustration of a neural networks. This structure is for illustration only and this includes one hidden layer, and two hidden units. Two input features, $X1$ and $X2$, work with the activation functions, $f1$ and $f2$, then obtain the outputs, $Y1$ and $Y2$.}
    	\label{fig:mlpc_structure}
\end{center}
\end{figure}

Multi-Layer Perceptron Classifier (MLPC) is a supervised artificial neural network with multiple hidden layers \citep{Rosenblatt1958, Fukushima1975, Fukushima1983}. Hidden layers which have several hidden units are invisible layers between input and output layer in neural networks, and are used to connect input features with each other. Each hidden unit is an activation function calculated by the product of weights and input. Using a neural network with one hidden layer as an example (Fig.~\ref{fig:mlpc_structure}), $X1$ and $X2$ are input features, $f1$ and $f2$ are the activation functions of hidden units calculated by (using $f1$ as an example) $f1=f\left( w0\cdot 1+w1X1+w2X2 \right) $, where $w$ are weights and $f$ represents an activation function as well. Through the calculation, it connects each input feature with hidden units by weights. Therefore, more hidden layers and more hidden units in each hidden layer can form more complicated connections of input features; however, the architecture with more hidden layers and hidden units is more time-consuming and can lead to overfitting problems. Similarly, the output layer also can be calculated from this concept. 

MLPC uses a back-propagation algorithm \citep{Werbos1974, Rumelhart1986}, which returns the error of predicted classification compared with the true label to the algorithm when the neural network is activated and the preliminary output is obtained. Algorithm adjusts the weights through the error until the error is lower than the tolerance which we set ${10}^{-5}$.  There are two hidden layers and 1,024 hidden units for each hidden layer in MLPC method we used. The learning rate is fixed to 0.001.
%figure
%SUBSECTION%
%figure
\begin{figure*}
\begin{center}
\graphicspath{}
	%MLPC structure introduction
	\includegraphics[width=1.7\columnwidth]{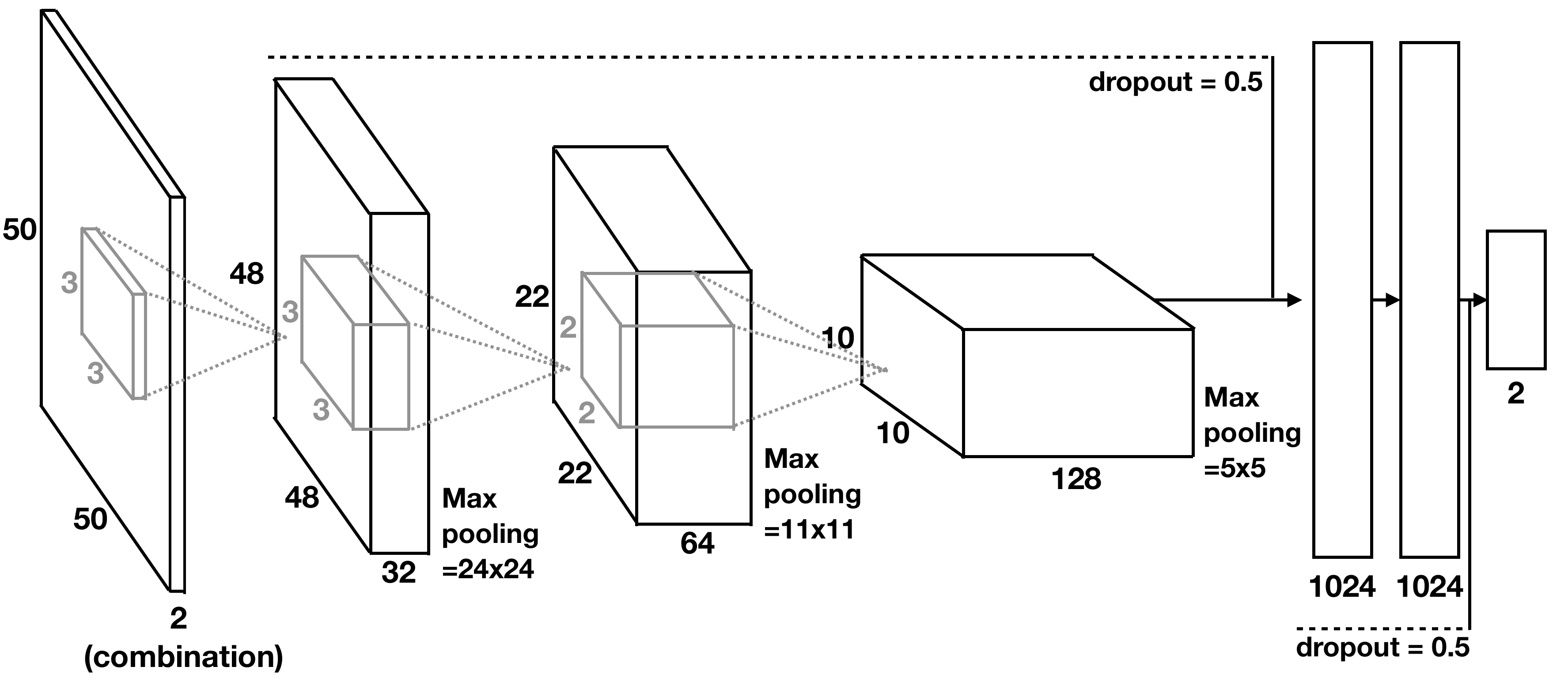}
   	\caption{The schematic overview of the architecture of CNN. The architecture starts from an input image with size 50 by 50 pixels, then three convolutional layers (filter: 32, 64, and 128). Each convolutional layer is followed a pooling layer. Two hidden layers with 1,024 hidden units for each are following the third convolutional layer. One dropout (p=0.5) follows after the third convolutional layer and the other follows after the second hidden layer. At last, there are two outputs in our CNN, `Ellipticals' and `Spirals'.}
    	\label{fig:cnn_architecture}
\end{center}
\end{figure*}
%paragraph
\subsection{Convolutional Neural Networks (CNN)}
Convolutional Neural Networks (CNN) started from the design of LeNet-5 \citep{LeCun1998}. However, CNN were not applied to the morphological classification of galaxies utill \citet{Dieleman2015} in the Galaxy Zoo Challenge\footnote{https://www.kaggle.com/c/galaxy-zoo-the-galaxy-challenge}. There are two main differences between artificial neural networks (e.g. MLPC) and CNN. One is that CNN has convolutional layers which are able to extract notable features from the input images by applying several filter matrices, and the other difference is the dimension of the input. 

Most machine learning algorithms are designed for dealing with 1D array input (e.g. parameter input), but some of them (e.g. SVM and neural networks) are able to deal with higher dimension data. However, the input still needs to be reshaped to 1D arrays for SVM and MLPC. On the contrast, CNN is designed for image input with three dimension arrays which means that in addition to the image itself, CNN has an extra dimension to store more information of image such as colours (RGB). 

Fig.~\ref{fig:cnn_architecture} shows the architecture of CNN that we use in this study. The input size of image is 50 by 50 pixels (Section~\ref{sec:stamp_creation}). We have 3 convolutional layers with filter sizes of 3, 3, 2, respectively, and each of them is followed with a pooling layer with size 2. These are then connected with two hidden layers with 1,024 hidden units for each layer. Additionally, two dropout layers are used to prevent overfitting, one follows the third convolutional layer (pooling layer), and the other comes after two hidden layers. The rectification of non-linearity is applied for each convolutional layer and hidden layer, and the softmax function is applied to the output layer to get the probability distribution of each type (all from the {\bf{Python}} package {\bf{lasagne.nonlinearities}}). We use {\it{Adam}} Optimiser, {\it{Nesterov momentum}}, and  set {\it{momentum=0.9}} according to \citet{Dieleman2015}, and the learning rate 0.001 and maximum 500 iterations for the CNN training.

%SECTION%
\section{Results}
\label{sec:results}
%figure
\begin{figure}
\begin{center}
\graphicspath{}
	%illusion of confusion matrix
	\includegraphics[width=0.6\columnwidth]{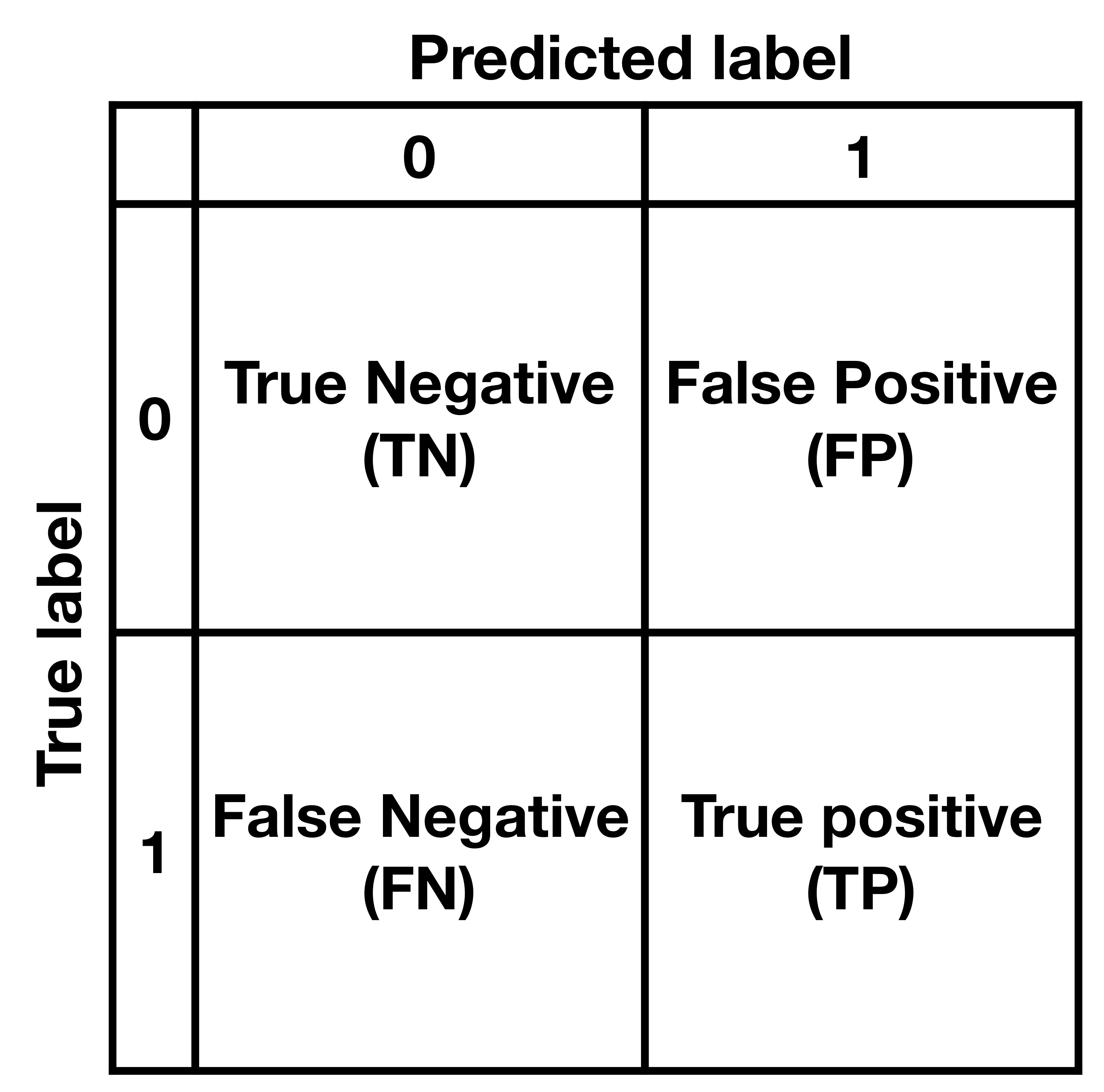}
   	\caption{The confusion matrix. The \textit{x}-axis label is the predicted label and the \textit{y}-axis label is the true label. The `0' means negative as well as Ellipticals type while `1' represents positive signal and Spirals type in this study.}
    	\label{fig:illusion_TP_FN}
\end{center}
\end{figure}
%SUBSECTION%
\subsection{The evaluation factors for models}
%Introduce how to test the model: ROC curve, the definition of True Positive Rate, False Positive Rate, Precision, Recall, and Accuracy.
%paragraph
We use the Receiver Operating Characteristic curve (ROC curve) \citep{Fawcett2006, Powers2011} to examine the performance of each method and dataset. On a ROC curve the \textit{y}-axis is the true positive rate and the \textit{x}-axis is the false positive rate; therefore, the closer the ROC curve gets to the corner (0,1), the better the performance is. The definition of true positive and the false positive are shown in Fig.~\ref{fig:illusion_TP_FN} in terms of the confusion matrix. Therefore, the true positive rate ($TPR$) and false positive rate ($FPR$) are defined as below,
%equation*
\begin{equation}
    	TPR=\frac { TP }{ TP+FN } ;\quad FPR=\frac { FP }{ FP+TN } .
\end{equation}
%paragraph
\noindent The definition of $TPR$ is identical to `recall ($R$)' in statistics which represents the completeness that shows how many true types have been picked, while `precision ($Prec$)' indicates the contamination which means how many picked types (predicted types) are true types. We are doing binary classification - positive: Spirals and negative: Ellipticals. Therefore, the recalls for Spirals and Ellipticals are shown below, 
%equation
\begin{equation}
    	Prec=\frac { TP }{ TP+FP };
\end{equation}
\begin{equation}
\label{eq:recall}
	R\left( 1 \right) =\frac { TP }{ TP+FN } ;\quad R\left( 0 \right) =\frac { TN }{ TN+FP }.
\end{equation}

%paragraph
Additionally, we also use the factor - the area under the ROC curve (AUC) as a performance evaluation for machine learning \citep{Bradley1997, Fawcett2006}. The meaning of AUC is the probability that a classifier ranks a randomly chosen positive example greater than a randomly chosen negative example. This factor also indicates the separability - how well the classifications can be correctly separated from each other. 

%SUBSECTION%
\subsection{The impact of rotated images}
\label{sec:rotated_images}
%figure
\begin{figure*}
\begin{center}
\graphicspath{}
	%roc curve for all methods we use in this study
	\includegraphics[width=2\columnwidth]{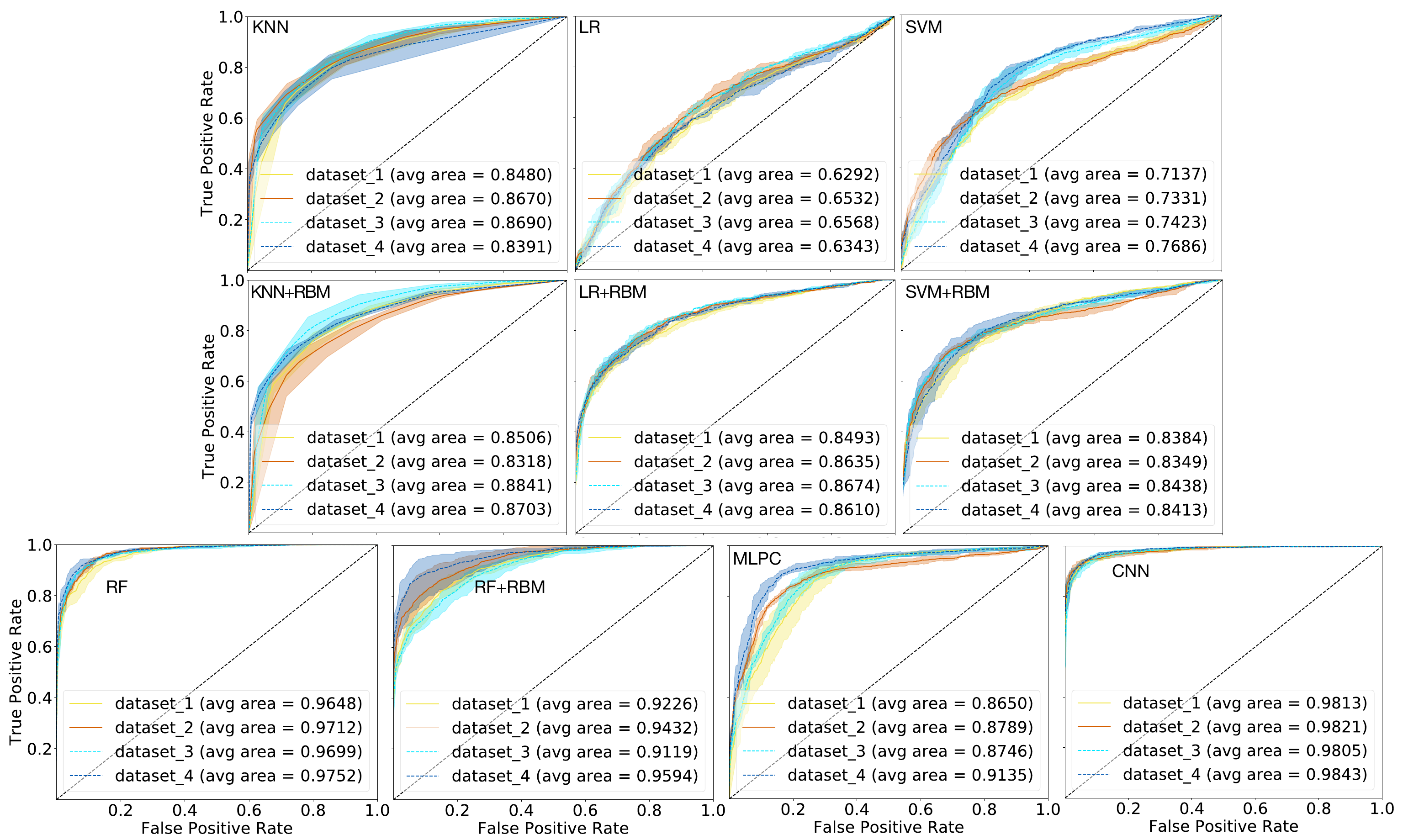}
   	\caption{The ROC curve of each method and each dataset using the raw images input (i). The abbreviation of the methods are the same as Table~\ref{tab:methods}. Different colours are for the different datasets (Table~\ref{tab:datasets}). Yellow, orange, cyan, blue are for dataset 1, 2, 3, 4, respectively. The lighter colour shading shows the scatters defined by the minimum and maximum of three reruns, and the lines inside are the averages of the three reruns. The black diagonal dashed line represents a random classification.}
    	\label{fig:roc_raw_minmax}
\end{center}
\end{figure*}
%paragraph
The ROC curves of each method and datasets are shown in Fig.~\ref{fig:roc_raw_minmax}. We show the results of raw images input (i) in this figure. Different colours represent different datasets such that the yellow, orange, cyan, blue lines represents datasets 1, 2, 3, 4, respectively (Table~\ref{tab:datasets}). The datasets 1 and 2 contain both the original images and the rotated images, and the datasets 3 and 4 only contain the rotated images. Meanwhile, the datasets 1 and 3 have an unbalance number of each type, conversely, the datasets 2 and 4 have an identical number for each classification. The lighter colour shadings are the scatters defined by the minimum and maximum over three reruns. The black diagonal dashed line indicates a random classification. 

First, the results of the LR and SVM methods, with and without combining with neural network, RBM show an improvement for LR and SVM when combining with RBM in Fig.~\ref{fig:roc_raw_minmax}. On the contrary, the performance of RF+RBM method shows slightly worse performance than the one of the RF method. Secondly, the scatters of the three reruns show small variance for each dataset, confirming the consistency of the reruns with each other. Additionally, as can be seen there are not large differences in the results between the different datasets. However, the slight shifts of the ROC curve occur within a few methods between the different datasets (e.g. MLPC). These are due to the slight differences in the total number of training samples for different datasets (Table~\ref{tab:datasets}). For example in MLPC, the dataset 4 has the maximum number of training data within the 4 datasets used ($\sim$12400 galaxies), so the performance of this dataset is the best in MLPC; the datasets 2 and 3 have very similar number of training data (the differences in number is only 67), thus they have a similar performance to each other. The dataset 1 has the least number of training data ($\sim$10400 galaxies), therefore, the performance is relatively worse. The shifts seen are also influenced by the condition of the balance between the ratio of each type (e.g. SVM and RF), for example, the datasets 1 and 3 are the unbalanced training data, so the shape of their ROC curve are similar to each other.  This is also the case for the datasets 2 and 4. To summarise, from Fig.~\ref{fig:roc_raw_minmax}, data augmentation through rotated images works fair to improve the performance of classification with machine learning.  

%SUBSECTION%
\subsection{Balance or Unbalance?}
\label{sec:balance}
%figure
\begin{figure}
\begin{center}
\graphicspath{}
	%\hspace*{-7ex}
	%recall for all methods and datasets
	\includegraphics[width=\columnwidth]{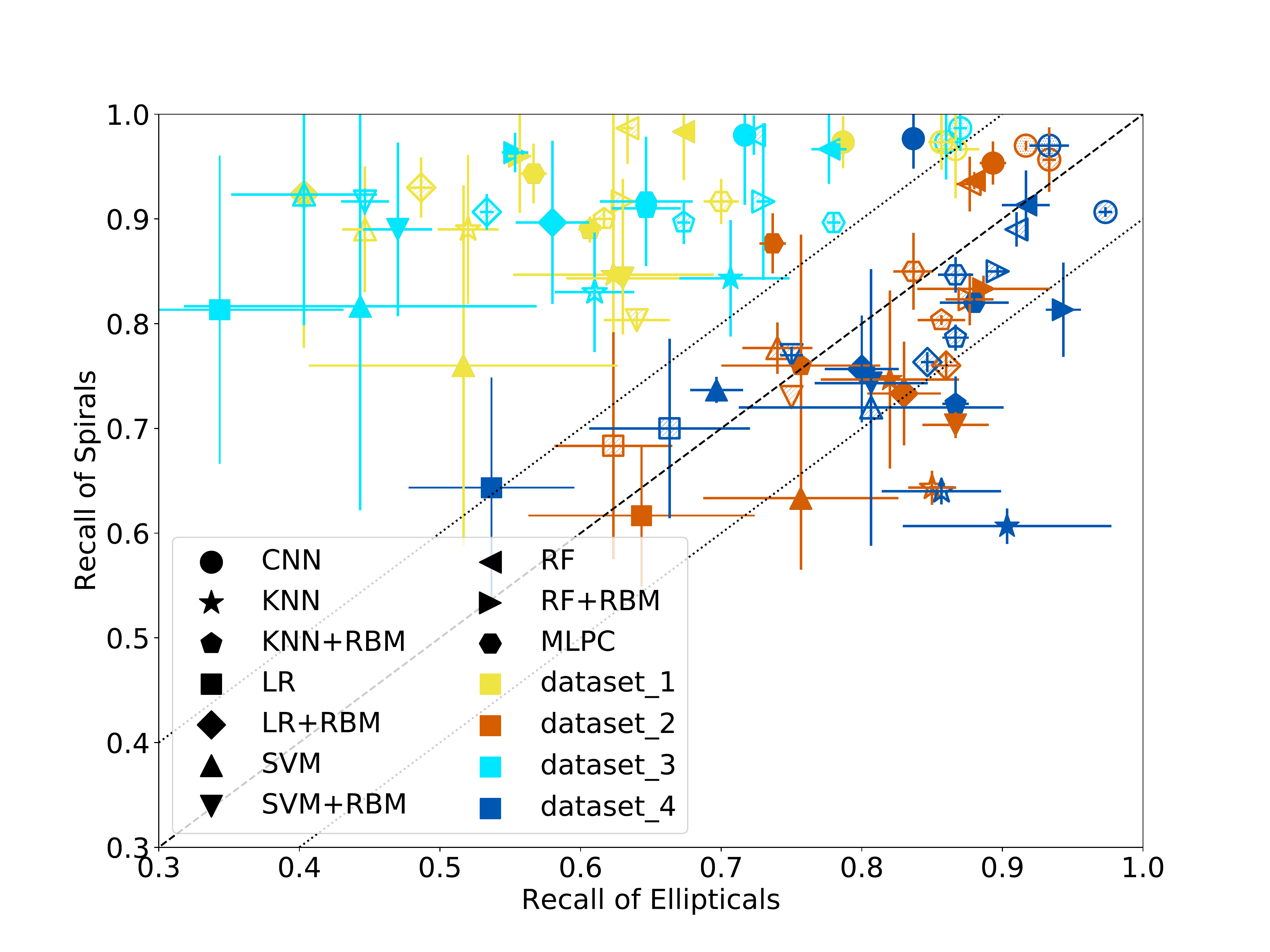}
   	\caption{The recalls of the Ellipticals and Spirals for all methods and the different types of the input data used. The colours represent the different datasets, while the different shape markers are the different methods. The different types of filled-points represent the different types of input. The fully-colour-filled markers are the raw images only (i), the diagonal-line-filled markers are the HOG images (ii), and those with dots are the combination input of the raw and HOG images (iii) which is only for CNN. The black dashed line represents the condition that $R(0)=R(1)$ (Equation~\ref{eq:recall}). The black dotted lines indicate that the differences in the recalls between these two types are within $\pm$0.1. The error bars are from the standard deviation of the three reruns.}
    	\label{fig:R_Prec}
\end{center}
\end{figure}
%paragraph
Here we investigate the influence of the balance between the number of each type in training data. Fig.~\ref{fig:R_Prec} shows the recalls of Ellipticals and Spirals for the different datasets using the different methods. The colour representation is the same as the ROC curve of Fig.~\ref{fig:roc_raw_minmax}, and the different methods are marked by the different shape markers. We obtain the value of the recall from equation~\ref{eq:recall} for Fig.~\ref{fig:R_Prec} by averaging the values from the three reruns. Different pattern types represent different types of input. The colour-filled points are the raw images input (i) while the points with diagonal-filled marker are the HOG images (ii), and with dotted-filled marker are the combination input (iii). The black diagonal dashed line shows the condition that $R(0)=R(1)$ (Equation~\ref{eq:recall}), and the black dotted lines show that the recall differences between these two types are within $\pm$0.1.

We observe that the unbalance training dataset 1 (yellow) and dataset 3 (cyan) are all above the upper dotted line which means that these two datasets generally have relatively higher recalls for Spirals compared to Ellipticals, and the differences of the recalls between Spirals and Ellipticals are larger than 0.1. For example, the result of the LR with the raw images input (i) (using the dataset 3 as an example shown as the leftmost cyan square in Fig.~\ref{fig:R_Prec}) has the recall of (0.34, 0.81) for Ellipticals and Spirals, respectively. We also observe that the LR, LR+RBM, SVM, and SVM+RBM methods have more seriously unbalanced results than other methods when using the unbalanced datasets (close to top-left in Fig.~\ref{fig:R_Prec}). This situation is due to the characteristics of these methods. For example, LR simply uses logistic functions to determine the decision boundary which can be easily shifted by unbalanced number of each type. On the other hand, \citet{Wu2003} discusses the skewed decision boundary of SVM caused by an unbalanced data such that the decision boundary is likely to be dominated by the support vector for the majority class.

On the other hand, most of the balanced dataset 2 (orange) and dataset 4 (blue) are located within two dotted lines which implies that these two datasets have similar recalls between Ellipticals and Spirals (the differences are smaller than 0.1). However, a few results of the balanced datasets in KNN have a higher recall of Ellipticals, but a relatively lower recall of Spirals (the orange and blue stars which are below the lower dotted line). KNN algorithm obtains the similarity between two images by calculating the `distance' between each pixel of two images (Section~\ref{sec:KNN}). Spirals have various shapes (e.g. different numbers of the spiral arms) while Ellipticals have a relatively simple appearance similar to one another. Therefore, it is easier for KNN to recognise Ellipticals than Spirals when we have the same numbers of both types within the training data.

We apply ten different common machine learning algorithms in this study and they show the consistent result in their balance except for KNN which we have discussed above; therefore, according to this discussion, the balance between the number of each type in training process is of great importance while using pixel input in most machine learning algorithms. In this figure, we also observe that the CNN method with a balanced datasets obtains the best recalls of both Ellipticals and Spirals.
%SUBSECTION%
\subsection{The effect of different types of input data}
\label{sec:different_input}
%figure
\begin{figure*}
\begin{center}
\graphicspath{}
	%roc curve for all methods we use in this study
	\includegraphics[width=2\columnwidth]{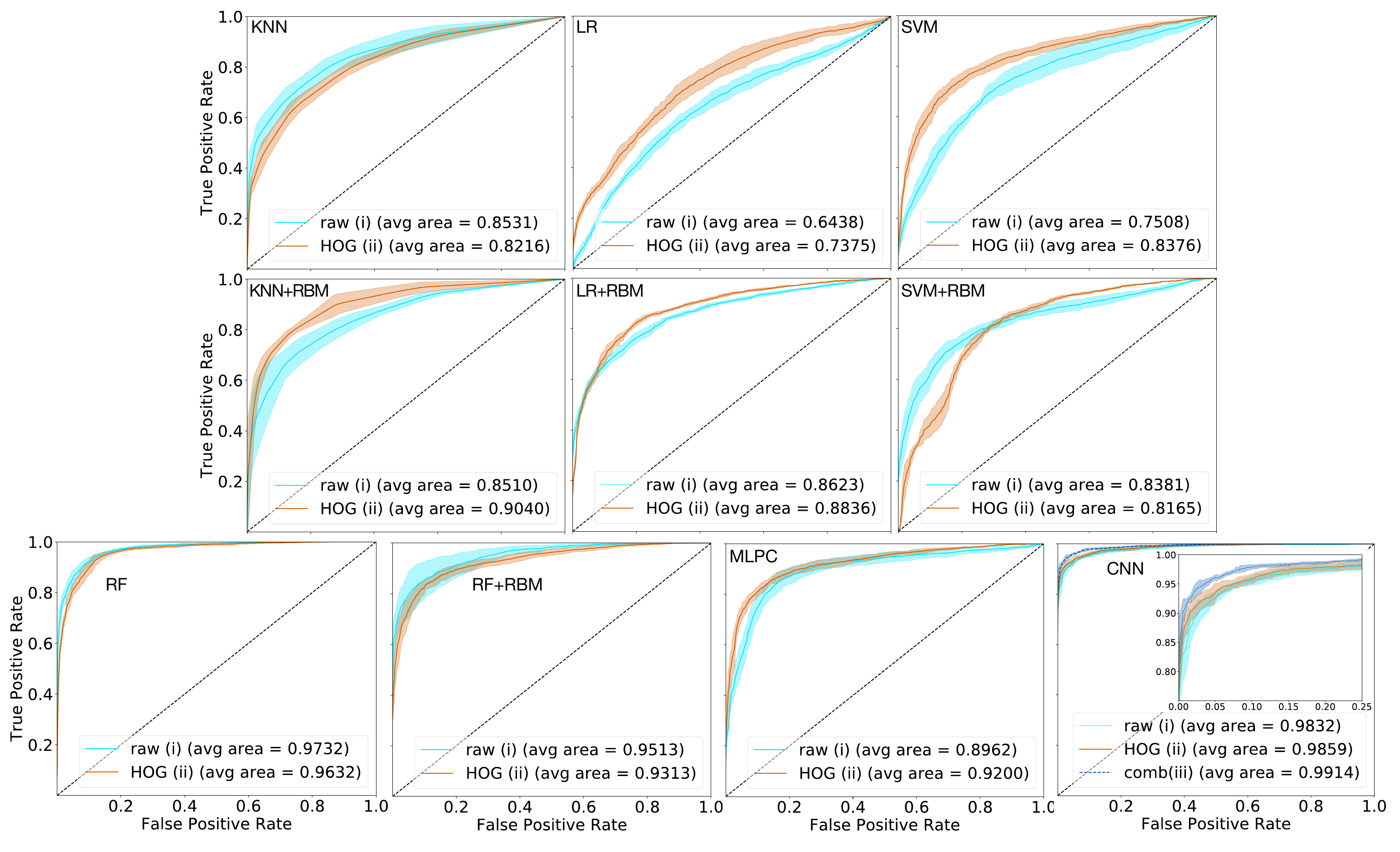}
   	\caption{The ROC curve for different types of input within each method. Different colours are for different input types of data. Cyan, orange, and blue are for raw images (i), HOG images (ii), and combination input (iii), respectively. The lighter colour shadings show the scatters defined by the standard deviation calculated through three runs of the balanced datasets 2 and 4. The lines inside the shading are the averages of the three reruns of the datasets 2 and 4. The black diagonal dashed line represents a random classification. The subplot in the CNN method is the zoom-in area from 0.75 to 1.0 in y-axis and from 0.0 to 0.25 in x-axis.}
    	\label{fig:roc_input_comp}
\end{center}
\end{figure*}
%figure
\begin{figure*}
\begin{center}
\graphicspath{}
	%accuracy of each dataset and method.
	\includegraphics[width=2\columnwidth]{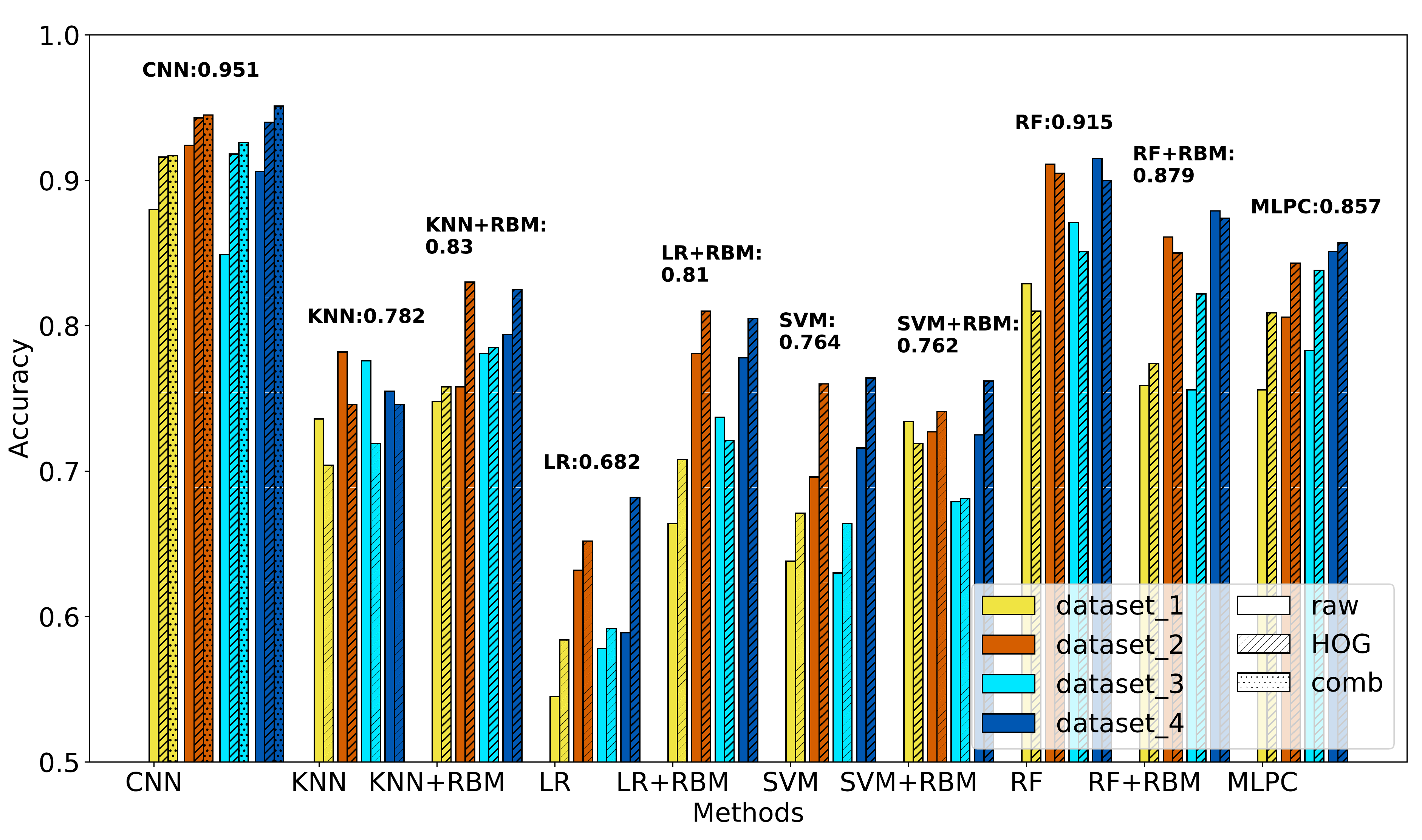}
   	\caption{The average accuracy (Equation~\ref{eq:accuracy}) of the three reruns versus each method with the different datasets and the different types of input shown. The y-axis is from 0.5 to 1.0. Colours represent different datasets such that yellow, orange, cyan, blue  represents dataset 1, 2, 3, 4 (Table~\ref{tab:datasets}), respectively. The different styles of shading are the different types of input data such that the fully-filled, the diagonal-line-filled, the dotted-filled represents the raw images (i), the HOG images (ii), and the combination input (iii), respectively. The labels above bars are the highest value of the accuracy for each method.}
    	\label{fig:accuracy}
\end{center}
\end{figure*}
%paragraph
Here we show the comparison results between the different types of input for each method (Fig.~\ref{fig:roc_input_comp}). We have 2 (3 for CNN) different types of input - the raw images (i) , the HOG images (ii), and the combinations input (iii) (for CNN only). Different colours in Fig.~\ref{fig:roc_input_comp} indicate different types of input such that cyan, orange, and blue are for the raw images (i), the HOG images (ii), and the combination input (iii), respectively. According to the discussions in section~\ref{sec:rotated_images} and section~\ref{sec:balance}, the results of the balanced datasets 2 and 4 are basically equivalent, and are better representations in our four datasets (Table~\ref{tab:datasets}). Therefore, we show the averages of the balanced datasets 2 and 4 after three reruns in Fig.~\ref{fig:roc_input_comp}, and the lighter colour shadings show the scatters defined by the standard deviation of the three reruns.

Fig.~\ref{fig:roc_input_comp} shows that the HOG images input successfully improves the performance in most of methods, except for KNN. Although the HOG image is able to extract the characteristics of the morphologies according to the value of the gradients, it also loses some of the detailed information  (i.e. the smaller fluctuations or gradients) and the smooth structure as well. Therefore, for KNN, the loss of the smooth structure in HOG images causes difficulties in determining the correct decision boundary. This result can be significantly improved by combining KNN with RBM when using the HOG images.

On the other hand, we observe that the application of HOG images shows an unapparent effect when combining RBM in LR+RBM, SVM+RBM and RF+RBM. We infer that this phenomenon is caused because that the RBM interlinks with the HOG features which have less information in the images than the raw images input. Therefore, it `annihilate' the effect of RBM and HOG which leave an unapparent change in these three methods. This effect is shown in both MLPC and CNN as well such that the HOG images input shows only a slight improvement in these two methods as well. However, increasing the number of hidden layers or more neurons in the neural networks helps to connect the HOG features with each other. Therefore, the improvements with HOG images in MLPC and CNN are qualitatively better than LR+RBM, SVM+RBM, and RF+RBM. A more qualitatively significant improvement is shown in CNN when we combine both the raw images input and the HOG images input (blue colour in CNN plot of Fig.~\ref{fig:roc_input_comp}).
%SUBSECTION%
\subsection{Comparison between methods}
\label{sec:comparison_bw_methods}
%paragraph
The definition of the accuracy used in Fig.~\ref{fig:accuracy} is shown below, 
%equation*
\begin{equation}
\label{eq:accuracy}
    	Accuracy=\frac { TP+TN }{ TP+FP+TN+FN }, 
\end{equation}
%paragraph
\noindent such the meaning of this is defined as how many successfully classified samples there are out of all the samples tested. The comparison of the accuracy for the different datasets and the different methods is shown in Fig.~\ref{fig:accuracy}. Through this figure we can observe the same situations as we have discussed in section~\ref{sec:different_input} such that most methods have a better performance when using the HOG images as input, except for the KNN where the HOG image input slightly reduces the performance, and the LR+RBM, SVM+RBM, and RF+RBM methods which the HOG images input gives no apparent improvement in performance. We also make another comparison of efficiency between all methods (Table~\ref{tab:efficiency}). Most methods were run on the 2.3GHz Intel Core i5 Processor with 16GB 2133 MHz LPDDR3 memory except for the `CNN (GPU)' which was run on the NVIDIA GeForce GTX 1080 Ti GPU.

Interestingly, the performance of RF wins the performance of MLPC with a faster computation time (Table~\ref{tab:efficiency}) using raw images which was totally unexpected. The further investigation for the capability of the RF on imaging data will be very helpful considering both the computing speed and a high accuracy the RF can reach. On the other hand, we can see that KNN and MLPC need less computation time but can reach a relatively good accuracy compared to other methods. Therefore, KNN and MLPC can be a good option when using pixel input. Additionally, although the KNN method has lower accuracy than MLPC, it applies raw images input which saves the preprocessing time that generates the HOG images (or other types of scaling).

The most successful methods when using pixel input in our study according to both the ROC curve (Fig.~\ref{fig:roc_input_comp}) and the comparison of accuracy (Fig.~\ref{fig:accuracy}) between each method is certainly CNN. Both of these two figures indicate that  the HOG image input helps the performance of CNN (Table~\ref{tab:comparison_input}).

Additionally, we create a new way to utilise the third dimension in CNN when we combine the raw image (i) with the HOG images (ii) which together we call a `combination input (iii)'. This shows a slight but qualitatively great improvement when using the combination input (iii) to do training in CNN (see CNN plot in Fig.~\ref{fig:roc_input_comp}). With the combination input (iii) and the balanced datasets, we can reach $\sim$0.95 accuracy with CNN using pixel input in this study (Table~\ref{tab:comparison_input}). 

On the other hand, \citet{Sreejith2018} proposes an `unanimous disagreement' indicating an object that all the classifiers agree with each other but disagree with the visual classification. In our study, we found only 3 galaxies out of 1,000 galaxies show an unanimous disagreement when considering all classifiers. These galaxies are all labelled as Spirals by the Galaxy Zoo 1 classification (GZ1) but classified as Ellipticals by our classifiers. We also visually confirmed that these galaxies are indeed Ellipticals. This unanimous disagreement is more likely caused by the debias process applied in GZ1 to statistically adjust the population of galaxies at a higher redshift rather than a simple visual misclassification.
%table
\begin{table}
	%the comparison between different type of input data for CNN dataset 2 and 4 only.
	\centering
	%\hspace*{-0.6cm}
	\begin{tabular}{rccl} 
		\hline
		\multicolumn{1}{c}{Methods} & {Training time} & {Testing time} & {accuracy} \\
		\hline
		\hline
		{KNN} & {$\sim$ 0.2 sec} & {$\sim$45 sec} & {0.782$\pm$0.027 (raw)}\\
		{KNN+RBM} & {$\sim$3000 sec} & {$\sim$45 sec} & {0.830$\pm$0.007 (HOG)}\\
		{LR} & {$\sim$7-8 sec} & {$\le$ 1 sec} & {0.682$\pm$0.040 (HOG)}\\
		{LR+RBM} & {$\sim$3000 sec} & {$\le$ 1 sec} & {0.810$\pm$0.012 (HOG)}\\
		{SVM} & {$\sim$800 sec} & {$\le$ 8 sec} & {0.764$\pm$0.029 (HOG)}\\
		{SVM+RBM} & {$\sim$3000 sec} & {$\le$ 8 sec} & {0.762$\pm$0.001 (HOG)}\\
		{RF} & {$\le$1 sec} & {$\le$ 5 sec} & {0.913$\pm$0.009 (raw)}\\
		{RF+RBM} & {$\sim$3000 sec} & {$\le$ 5 sec} & {0.870$\pm$0.031 (raw)}\\
		{MLPC} & {$\sim$18 sec} & {$\le$ 3 sec} & {0.857$\pm$0.010 (HOG)}\\
		{CNN} & {$\sim$3000 sec} & {$\le$ 5 sec} & {0.951$\pm$0.005 (comb)}\\
		{CNN (GPU)} & {$\sim$360 sec} & {$\le$ 5 sec} & {0.951$\pm$0.005 (comb)}\\
		\hline
		\hline
	\end{tabular}
	\caption{The comparison of the computing time (per $\sim$1000 galaxies) for each method. The `accuracy' is the best accuracy shown in Fig.~\ref{fig:accuracy}. The first ten methods were run on  the 2.3GHz Intel Core i5 Processor with 16GB 2133 MHz LPDDR3 memory, while the sixth method `CNN (GPU) was run on the NVIDIA GeForce GTX 1080 Ti GPU.}
	\label{tab:efficiency}
\end{table}
%table
\begin{table}
	%the comparison between different type of input data for CNN dataset 2 and 4 only.
	\centering
	\begin{tabular}{ccc} 
		\hline
		\multicolumn{1}{c}{Input Types} & {accuracy} & {$R_{ 01 }$} \\
		\hline
		\hline
		\multirow{2}{*}{raw (i)} & {dataset 2: 0.924$\pm$0.013} & {0.933}\\
		& {dataset 4: 0.906$\pm$0.018} & {0.907}\\
		\hline
		\multirow{2}{*}{HOG (ii)} & {dataset 2: 0.943$\pm$0.016} & {0.940}\\
		& {dataset 4: 0.940$\pm$0.003} & {0.940}\\
		\hline
		\multirow{2}{*}{comb (iii)} & {dataset 2: 0.945$\pm$0.004} & {0.947}\\
		& {dataset 4: 0.951$\pm$0.005} & {0.953}\\
		\hline
		\hline
	\end{tabular}
	\caption{The comparison between the different types of input in CNN when using the datasets 2 and 4 (Table~\ref{tab:datasets}). The total number of testing images is 1,000 galaxies. The definition of the accuracy is according to Equation~\ref{eq:accuracy}. The value of $R_{ 01 }$ is the recall value of Ellipticals and Spiral (Eqaution~\ref{eq:recall}) after taking a weighted average, and the value of this is shown in the table as the three reruns average of $R_{ 01 }$.}
	\label{tab:comparison_input}
\end{table}
%SECTION%

\section{Further Discussion}
\label{sec:discussion}

We have already discussed some of our results in Section~\ref{sec:results} while presenting the results. In the last section we concluded that the best method of these ten supervised machine learning methods is Convolutional Neural Networks (CNN), the further analysis and the discussion of CNN is essential for all future usage (Section~\ref{sec:analysis_of_cnn}), as well as the investigation of misclassification and galaxies with low predicted probabilities (Section~\ref{sec:analysis_of_failure}).
%SUBSECTION%

\subsection{Analysis of Convolutional Neural Network (CNN)} 
\label{sec:analysis_of_cnn}
Here we discuss in more detail the results of our CNN machine learning classification.  We use a default criterion for the classification in CNN such that  the probability $(p)>0.5$ is the criterion for classification;  namely, Ellipticals or Spirals with $p>0.5$ will be classified as that type. When we change the criterion to $p \ge 0.8$, namely, any types with $p\ge0.8$ are classified as the predicted type, and if both types have $p<0.8$ then that galaxy will be classified as `Uncertain type'. With this criterion, we separate our testing data into three different classes: Ellipticals, Spirals, and Uncertain.   Using the combination input (iii), the accuracy of classification increases to $\sim$0.97 (Table~\ref{tab:criterion_clf}).
%table
\begin{table}
	%the results of dataset 2 and dataset 4 with criterion p=0.8
	\centering
	\begin{tabular}{cccccc} % four columns, alignment for each
		\hline
		{} & {accuracy} & {$R_{ 01 }$} & {${ N }_{ \text{classifiable} }$} & ${ N }_{ \text{uncetain} }$\\
		\hline
		\hline
		{dataset 2} & {0.974$\pm$0.004} & {0.973} & {912} & {88} \\
		{dataset 4} & {0.974$\pm$0.003} & {0.973} & {927} & {73} \\
		{Maximum} & {0.987$\pm$0.001} & {0.99} & {958} & {42} \\
		\hline
		\hline
		
	\end{tabular}
	\caption{The average result of the classification success with the classification criterion $p > 0.8$ through using CNN for dataset 2, dataset 4 (Table~\ref{tab:datasets}), and the result of the maximum available number of training data in our study with the combination input (iii) which includes both raw and HOG images. The total number of testing galaxies is 1,000. The definition of accuracy (Equation~\ref{eq:accuracy}) and the meaning of $R_{ 01 }$ are same as in Table~\ref{tab:comparison_input}. ${ N }_{ \text{classifiable} }$ and ${ N }_{ \text{uncertain} }$ are the number of testing data which are classifiable (namely $p\ge0.8$) and uncertain (probabilities of both types $(p)<0.8$), respectively.}
	\label{tab:criterion_clf}
\end{table}

%paragraph
%Another thing worth mentioning is that due to the limited number of rotation angles we use, we are able to increase our number of data by up to a factor of 35; therefore, according to the total number of training data  we have ($\sim$12,000, the average of dataset 2 and 4), we would only need $\sim$170 galaxies for each type to reach $\sim$0.97 accuracy for binary classification of Ellipticals and Spirals with a combination of  input (iii) and using a criterion $p=0.8$ within our CNN architecture.

Secondly, increasing the number of training samples should intuitively  improve the performance; however, we investigate whether this assumption is correct.   We increase the number of our training samples by the rotated images, and keep the balance between the number of both types of galaxies. The maximum balanced number of the training data in our study is 53,663 (S: 26,839; E: 26,824).

\begin{figure}
\begin{center}
\graphicspath{}
	%accuracy of each dataset and method.
	\includegraphics[width=\columnwidth]{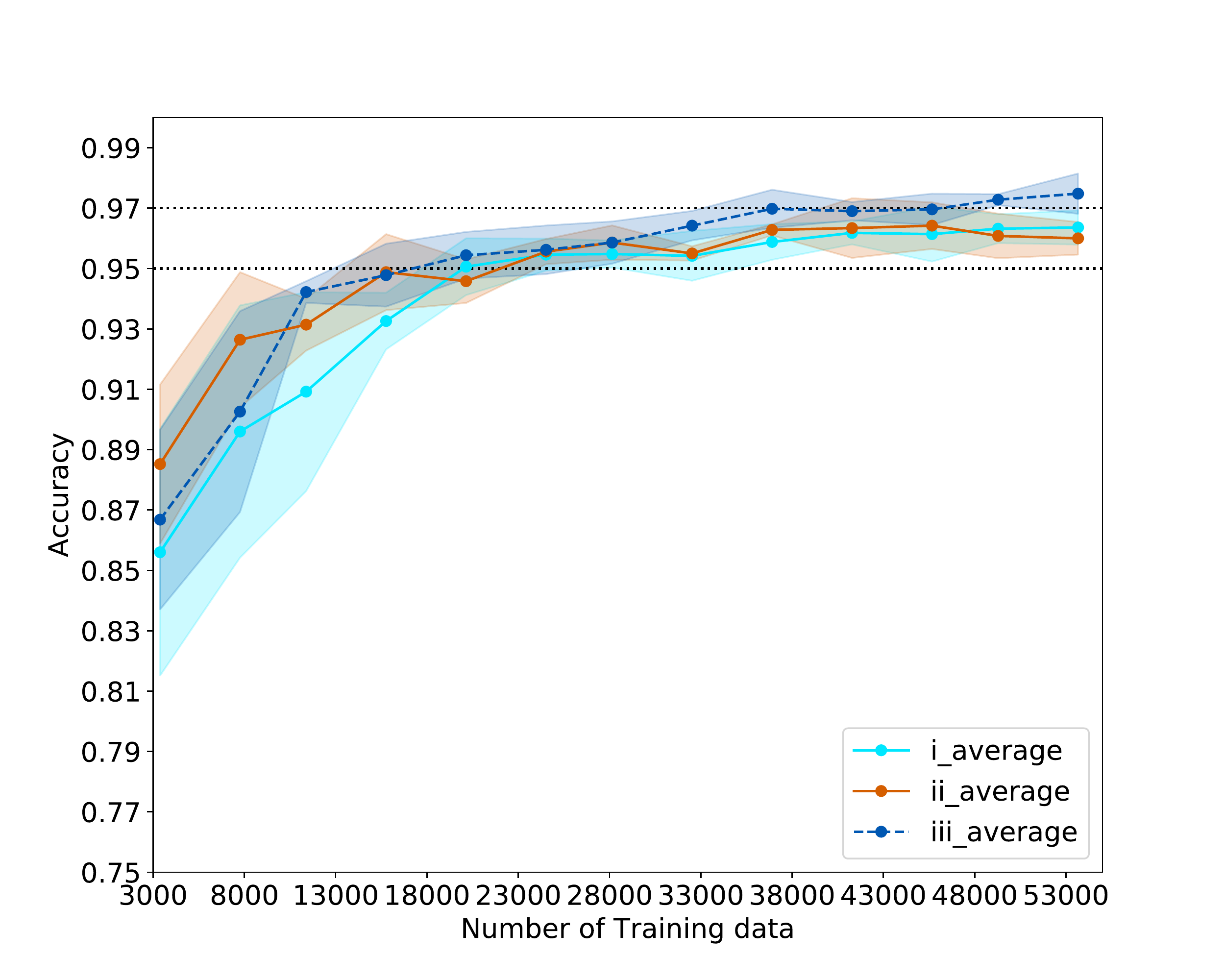}
   	\caption{The accuracy versus the number of training data with different types of input. Different colours show different types of input such that cyan, orange, blue are for the raw images (i), the HOG images (ii), and the combination input (iii), respectively. The lighter colour areas show the scatters of the standard deviation calculated by the five reruns, and the lines inside shadings show the average of the five reruns. The two dotted horizontal lines indicate the accuracy of 0.95 and 0.97.}
    	\label{fig:accuracy_trainN}
\end{center}
\end{figure}

In Fig.~\ref{fig:accuracy_trainN}, we observe that the increased rate of accuracy remains basically positive, but this decreases as the number of training data increases.   This shows that there is likely a maximum accuracy limitation within the CNN method for galaxy classification. This indicates that our combination input (iii) has a better performance than the other two types of input data as we increase the number of training data, and the combination input (iii) is the only one which is able to reach over the accuracy of $\sim$0.97 without any condition.

Therefore, we apply our maximum number of training data (53,663) with the combination input (iii) to do the training, and combine it with the classification criterion $p=0.8$. We then obtain a high accuracy of $\sim$0.987 in the morphological classification of galaxies. The result is shown in the third row of Table~\ref{tab:criterion_clf}.
%figure

%SUBSECTION%
\subsection{Origin of Classification Failures}
\label{sec:analysis_of_failure}
As shown in the above section, we are able to reach a high classification accuracy of $\sim$0.987 by using CNN with the maximum number of the training data with a combination of input (iii), and the criterion of the probability $p\ge0.8$. However, the $<100$ percent accuracy indicates that there are a few galaxies misclassified but with high predicted probabilities ($p\ge0.8$). On the other hand, there are also a few galaxies ($\sim$42 out of 1,000 testing galaxies) which are non-classifiable (lower predicted probability $p<0.8$ in both Ellipticals and Spirals). Table~\ref{tab:prob_clf} shows the fraction of the samples within a  range of probability (out of 1,000 testing galaxies), and the number of misclassification out of the galaxies within a probability range. It indicates that the classifications with higher probabilities ($p\ge0.8$) are much less often misclassified. However, it also shows that galaxies with the predicted probabilities between 0.7-0.8 have a  higher misclassified rate than the predicted probabilities  between 0.6-0.7. This means that there are some galaxies with relatively higher predicted probabilities but which are misclassified by our CNN.
%table
\begin{table}
	%the results of dataset 2 and dataset 4 with criterion p=0.8
	\centering
	\begin{tabular}{ccc} % four columns, alignment for each
		\hline
		{probability} & {sample fraction} & {misclassification} \\
		\hline
		\hline
		{$p\ge0.8$} & {0.958} & {0.0142}\\
		{$0.7\le p<0.8$} & {0.0184} & {0.239}\\
		{$0.6\le p<0.7$} & {0.0302} & {0.132}\\
		{$0.5\le p<0.6$} & {0.0114} & {0.368}\\
		\hline
		\hline
		
	\end{tabular}
	\caption{The fraction of the samples out of 1000 testing galaxies, and the fraction of misclassification within a certain probability range calculated by being divided by the sample number. The results are the average of five reruns.}
	\label{tab:prob_clf}
\end{table}

In this section, we define two types of failures by our CNN. One is the misclassification with the comparison to the Galaxy Zoo 1 classification with high predicted probabilities ($p\ge 0.8$), that are galaxies which were classified with high probabilities with CNN but which later turned out to have a different classification in Galaxy Zoo.  The other type of `failed' classification are those galaxies with low predicted probabilities ($p<0.8$ in both types) of being either elliptical or spiral.  We investigate the origin of these `failures' in this section.

%SUB-SUBSECTION%
\subsubsection{The failure with high probability: The misclassification of the classifiable galaxies}
\label{sec:highP_failure}

\begin{figure}
\begin{center}
\graphicspath{}
	%accuracy of each dataset and method.
	\includegraphics[width=\columnwidth]{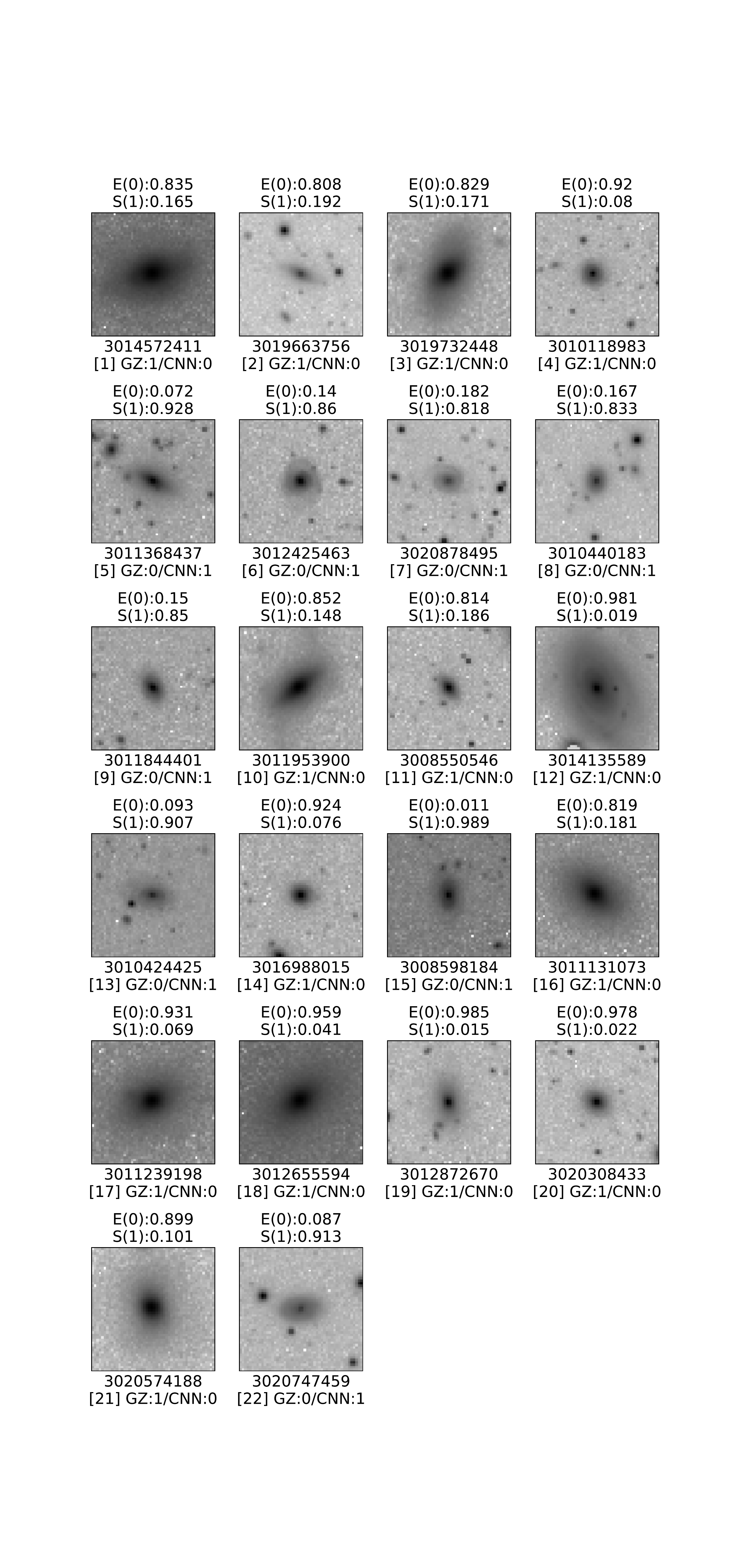}
   	\caption{The misclassified galaxies with high probabilities ($p\ge0.8$) comparing the classification of Galaxy Zoo 1 and our CNN. On the top of the images shows the probabilities of being Ellipticals, E(0) and Spirals, S(1) by our CNN. The line below the image shows the ID number of the galaxies in Dark Energy Survey (DES), and the second row shows the classifications by Galaxy Zoo and our CNN.}
    	\label{fig:highP_failure_log}
\end{center}
\end{figure}

We rerun five times the best combination of our method (i.e. the CNN trained by the maximum balanced number of training data and the combination input (iii), and classified by the criterion $p=0.8$), and we then collect all the misclassification of the classifiable galaxies from these five reruns together, obtaining 22 galaxies in total (Fig.~\ref{fig:highP_failure_log}).  Misclassification in this sense is that what we get from our CNN analysis differs from the Galaxy Zoo classification.  Most of these 22 galaxies are repeatedly misclassified between these five reruns, in Fig.~\ref{fig:highP_failure_log}, objects 1-7 only show up once, objects 8-17 are repeated more than twice, and objects 18-22 are repeatedly showing up in five reruns. There are two main probable reasons for these misclassifications with a high probability through our CNN method.  One is that we use the galaxy images with linear scale (including HOG images) on our CNN training, so in some cases, even if it shows the feature of Spirals in logarithmic scale, it is just a point source, a round object, or a large bright area in linear scale. Therefore, they prefer to be classified as Ellipticals rather than Spirals in our CNN. This will be further discussed in the section~\ref{sec:log}.

\begin{figure*}
\begin{center}
\graphicspath{}
	%accuracy of each dataset and method.
	\includegraphics[width=1.5\columnwidth]{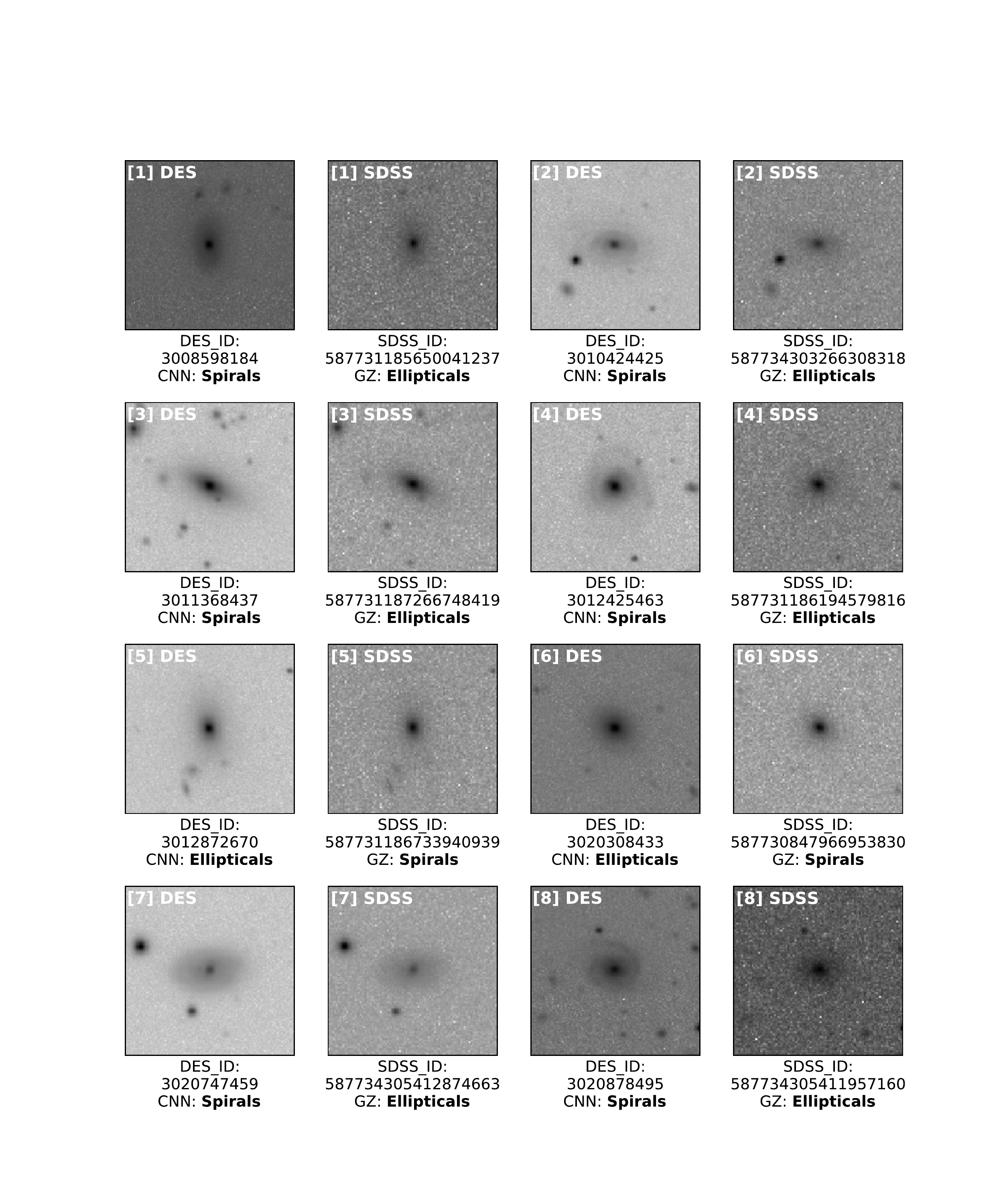}
   	\caption{Examples of the incorrect label from GZ1 with SDSS imaging. The figures under each number show the galaxy images of DES and SDSS, and their ID numbers. The label of `CNN' shows the predicted label from our method, and which of 'GZ' shows the label from the Galaxy Zoo 1 catalogue.}
    	\label{fig:GZ1_comp}
\end{center}
\end{figure*}

The other reason for the differences is due to misclassifications by the Galaxy Zoo 1 (GZ1). We apply visual classifications which have over 80$\%$  agreement between volunteer classifiers in the GZ1 catalogue in which we use to label our DES data. When we compare the SDSS imaging to the DES imaging, we can see some GZ1 classifications based on the SDSS data were simply wrong. Some examples are shown in Fig.~\ref{fig:GZ1_comp}. Most of them are revealed to be misclassifications due to the better resolution and deeper depth of the DES data than the  SDSS data. With higher resolution of the DES data, we reveal more detailed structure than the SDSS data (e.g the number 4 and 8 in Fig.~\ref{fig:GZ1_comp} which show clear spiral structures in the DES data but nothing in the SDSS data). We will further discuss this in Section~\ref{sec:misclf_GZ}.

On the other hand, we also discover that some galaxies with large, bright, and oval structure are easy to misclassify using our method. These galaxies are lenticular galaxies when examined on the DES imaging. The main reason for their misclassifications is because there is not a class for lenticular galaxies in the Galaxy Zoo project. Lenticular galaxy is difficult to see by visual classification and typically requires high resolution and deep imaging, even for nearby galaxies. Some of them are therefore classified as Spirals, and some of them are recognised as Ellipticals in the GZ1 catalogue. The details will be discussed in the next section (Section~\ref{sec:uncertain}) as most of these galaxies generally have lower predicted probabilities of being either elliptical or spiral.
%figure
%figure
%\begin{figure}
%\begin{center}
%\graphicspath{{figure/}}
	%accuracy of each dataset and method.
%	\includegraphics[width=\columnwidth]{highP_failure_linear_2.pdf}
%   	\caption{The misclassified galaxies with high probability ($p\ge0.8$) in linear scale.}
%    	\label{fig:highP_failure_linear}
%\end{center}
%\end{figure}
%figure
%\begin{figure}
%\begin{center}
%\graphicspath{{figure/}}
	%accuracy of each dataset and method.
%	\includegraphics[width=\columnwidth]{highP_failure_HOG_2.pdf}
%  	\caption{The HOG images of the misclassified galaxies with high probability ($p\ge0.8$).}
%    	\label{fig:highP_failure_HOG}
%\end{center}
%\end{figure}
%figure
%SUB-SUBSECTION%
\subsubsection{The failures at low probability: Uncertain type}
\label{sec:uncertain}

In this section, we investigate the galaxies with lower predicted probabilities ($p<0.8$) for classification as either elliptical or spiral in the five reruns of our best method. The majority of the samples with lower probabilities are repeated between five reruns, and some of them also show up in the previous section (Section~\ref{sec:highP_failure}) which are misclassified but with high probabilities. The probabilities of these galaxies vary significantly between each rerun.

\begin{figure}
\begin{center}
\graphicspath{}
	%accuracy of each dataset and method.
	\includegraphics[width=\columnwidth]{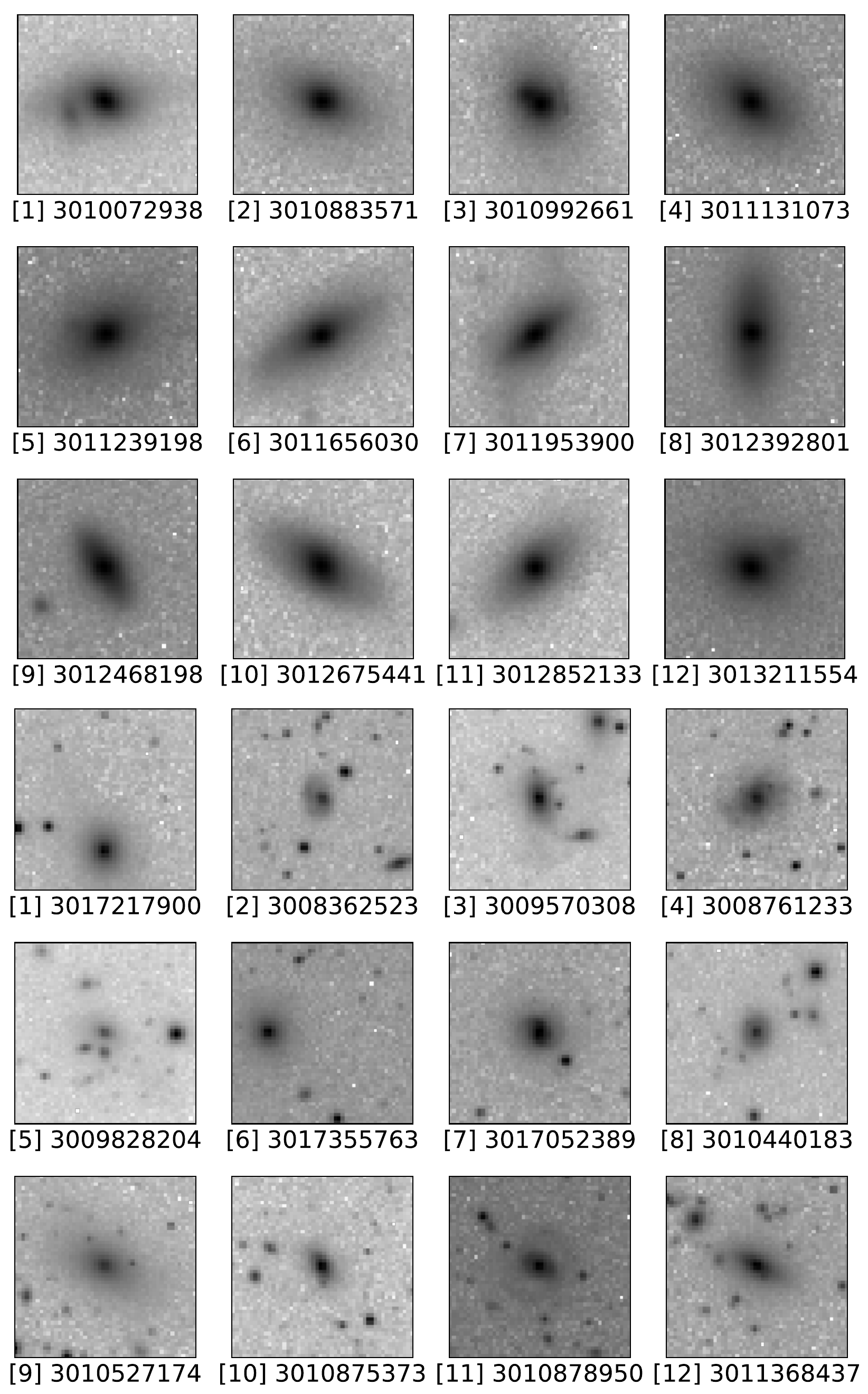}
  	\caption{Examples of the galaxies with low probabilities of classification as either spiral or elliptical. {\it Top 1-12:} these objects are turned out to be lenticular galaxies (S0) in cluster inspection. {\it Bottom 1-12:} the other types of galaxies.}
    	\label{fig:uncertain}
\end{center}
\end{figure}

The appearance of these galaxies can be separated into two types. One type are the galaxies which look large, oval, and bright ({\it Top 1-12} in Fig.~\ref{fig:uncertain}), and the other type are those which do not appear this way, e.g. galaxies which are relatively fainter or with large bulge and spiral structure at the same time, or the target galaxy is shifted significantly away from the centre of the image ({\it Bottom 1-12} in Fig.~\ref{fig:uncertain}). 

The galaxies with large and oval structure are lenticular galaxies which we discussed in the previous section (Section~\ref{sec:highP_failure}). As discussed there is not a  lenticular galaxy class in the GZ project, nor can these types be easily seen in SDSS data, therefore, the classification of these galaxies in the GZ1 catalogue are such that half of them are classified as Spirals, and half of them are classified as Ellipticals. Because lentinculars are neither spirals or ellipticals, their structure confuses our CNN such that it gives lower probabilities for these galaxies to be of either type.  This is a 'rediscovery' of lenticulars, and shows the power of machine learning for discovering new types of galaxies, as we did not expect this to occur.  

%figure
%\begin{figure}
%\begin{center}
%\graphicspath{{figure/}}
	%accuracy of each dataset and method.
%	\includegraphics[width=\columnwidth]{example_multi_objects_r.pdf}
%  	\caption{Examples of the galaxies with lower probabilities surrounded by multiple objects or bright points.}
%    	\label{fig:multi_objects}
%\end{center}
%\end{figure}
%figure
%SUB-SUBSECTION%
\subsubsection{Combined with logarithmic scale images}
\label{sec:log}
According to the discussion in the section~\ref{sec:highP_failure}, we investigate the impact on our classification with CNN when using images with logarithmic scale (hereafter, log images) to train our CNN algorithm by using the datasets 2 and 4 (Table~\ref{tab:datasets}). In addition to the log images, we also combine the log images with our combination input (iii) as the input to train our CNN. The comparison of the results are shown in Table~\ref{tab:log_comp}. 

Comparing Table~\ref{tab:log_comp} with Table~\ref{tab:comparison_input} shows a significant improvement when using the log images, and the combination of the log images and our combination input (iii) shows a better accuracy than just using the log images as input. 

However, comparing Table~\ref{tab:log_comp} with Table~\ref{tab:criterion_clf} shows that there are not significant differences in the performance from log images input to the other three types of input, (i), (ii), (iii), when we train our CNN through the maximum available number of the training data. This means that there is an intrinsic limitation of our method. This limitation can also be seen in Fig.~\ref{fig:accuracy_trainN} in Section~\ref{sec:analysis_of_cnn}. 

Therefore, we conclude that although adding the log images as input helps the performance, it still has no apparent difference from our result when we apply the maximum number of training data to our CNN.
%table
\begin{table}
	%the results of dataset 2 and dataset 4 with criterion p=0.8
	\centering
	%\hspace*{-0.5cm}
	\begin{tabular}{ccccc} % four columns, alignment for each
		\hline
		{} & {} & {} & {comb. input(iii)} \\
		{} & {log image} & {} & {+log image}\\
		\hline
		\hline
		{} & {accuracy} & {$R_{ 01 }$} & {accuracy} & {$R_{ 01 }$} \\
		\hline
		\hline
		{dataset 2} & {0.950$\pm$0.006} & {0.947} & {0.952$\pm$0.006} & {0.950} \\
		{dataset 4} & {0.954$\pm$0.004} & {0.953} & {0.964$\pm$0.007} & {0.967} \\
		
		{Maximum} & {0.973$\pm$0.002} & {0.970} & {0.971$\pm$0.005} & {0.973} \\
		{Max ($p=0.8$)} & {0.987$\pm$0.004} & {0.987} & {0.987$\pm$0.003} & {0.987} \\
		\hline
		\hline
		
	\end{tabular}
	\caption{The comparison of the accuracy (Equation~\ref{eq:accuracy}) and the recalls (Same as Table~\ref{tab:comparison_input}) between the inputs of the log images and the combination of log images and combination input (iii) by using the dataset 2, dataset 4 (Table~\ref{tab:datasets}), and the maximum number of training data.}
	\label{tab:log_comp}
\end{table}
%SUB-SUBSECTION%
\subsubsection{The advantage of Dark Energy images and the misclassifications by Galaxy Zoo project}
\label{sec:misclf_GZ}

We have discussed the incorrect labels by Galaxy Zoo in previous sections. As discussed, the main reason to reveal the misclassification by SDSS imaging Galaxy Zoo is because of the better resolution (${ 0. }^{ \prime \prime  }263$ per pixel) and deeper depth of DES data ($i=22.51$) \citep{Abbott2018}.

These wrong labels not only influence the results of our CNN, but also contaminate the training set. Therefore, we remove the potential misclassified galaxies from the training set. We purify our training set by excluding the suspected misclassified galaxies then use the criteria shown in Table~\ref{tab:confirm_criterion} to confirm or dismiss our suspected misclassifications. We then rerun our CNN classification five times on each new training set and obtain five new CNN models on the new classifications.  After carrying out this purification twice, and then retraining and updating our list of suspects, we obtain two lists of these galaxies: one is the confirmed misclassified galaxies by the Galaxy Zoo, and the other are the suspected misclassified galaxies. 

\begin{figure*}
\begin{center}
\graphicspath{}
	%accuracy of each dataset and method.
	\includegraphics[width=1.95\columnwidth]{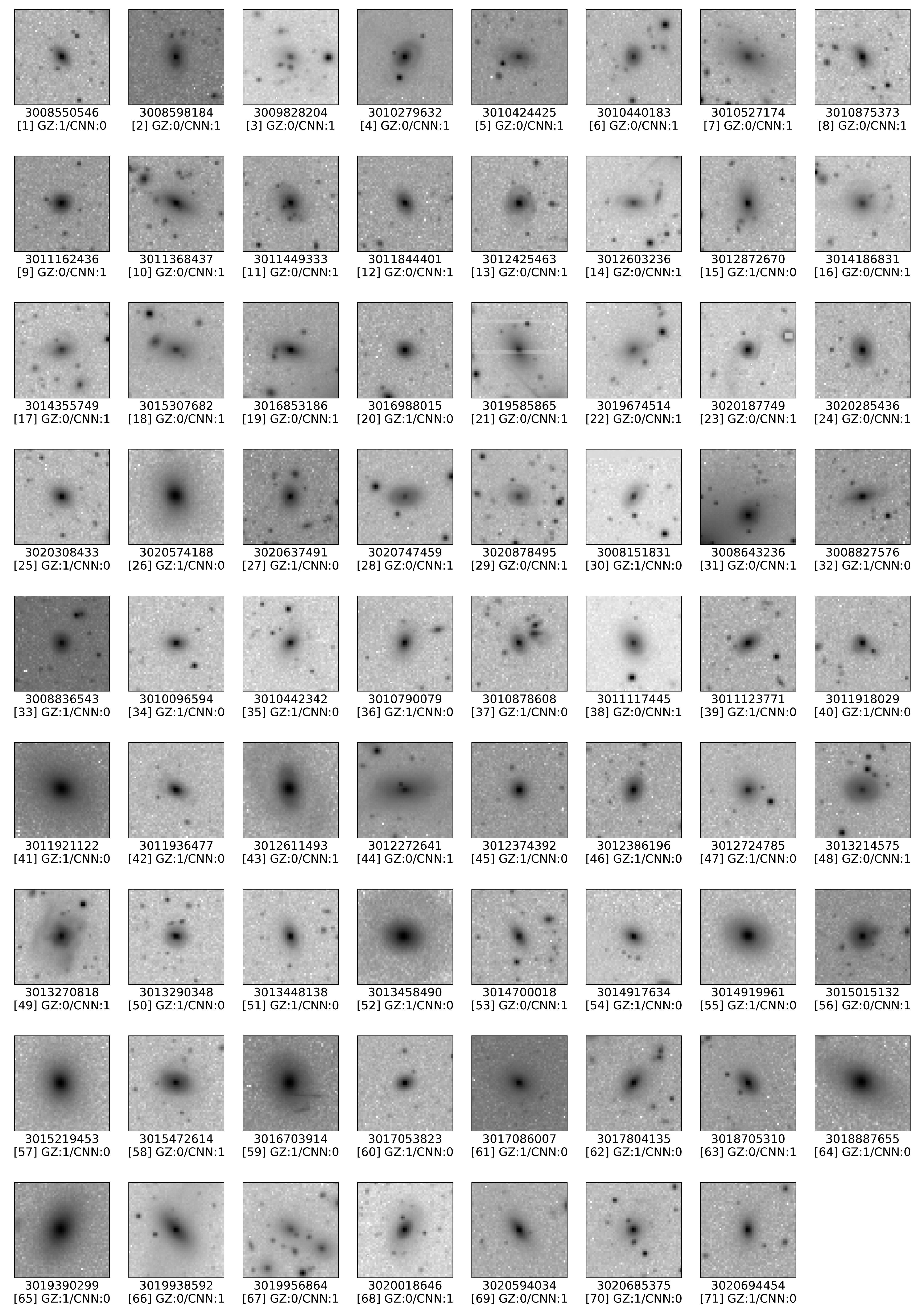}
  	\caption{The confirmed list of the misclassified galaxies in the Galaxy Zoo 1 catalogue. The first row underneath the images is the ID numbers of galaxies, and the second row shows the classification by Galaxy Zoo (GZ) and our CNN (CNN).}
    	\label{fig:confirmed}
\end{center}
\end{figure*}

\begin{figure*}
\begin{center}
\graphicspath{}
	%accuracy of each dataset and method.
	\includegraphics[width=1.95\columnwidth]{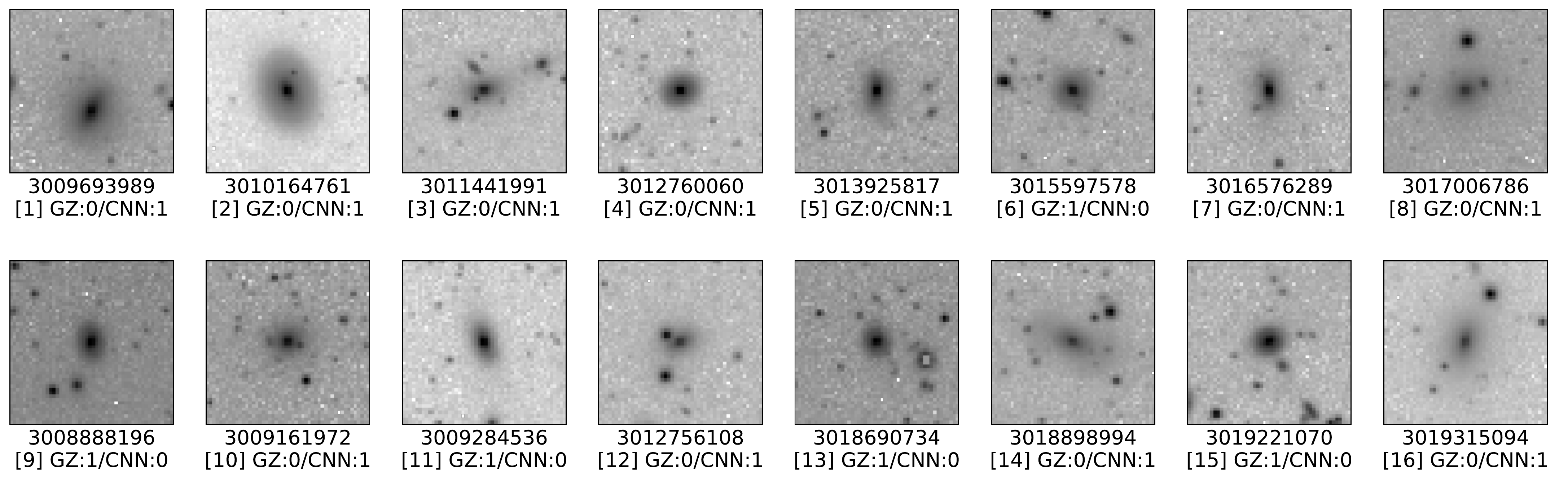}
  	\caption{The suspected list of the misclassified galaxies in the Galaxy Zoo 1 catalogue. The first row underneath the images is the ID numbers of galaxies, and the second row shows the classification by Galaxy Zoo (GZ) and our CNN (CNN).}
    	\label{fig:less_confirmed}
\end{center}
\end{figure*}

The images of these systems are shown in Fig.~\ref{fig:confirmed} and Fig.~\ref{fig:less_confirmed}. There are $\sim2.5\%$ misclassified galaxies in the Galaxy Zoo 1 catalogue out of ~2,800 in our study as revealed by using DES images and our CNN, and $\sim0.56\%$ are suspected candidates in our study. We then correct our training set according to these two lists. We change the label of the confirmed misclassified galaxies, and exclude the suspected misclassified galaxies from the training set, then do the training with the maximum available number which is 53,141 galaxies in total (E: 26,344; S: 26,797). We then change the label of the confirmed misclassified galaxies in the testing set as well. 
%table
\begin{table}
	%the results of dataset 2 and dataset 4 with criterion p=0.8
	\centering
	\begin{tabular}{cl} % four columns, alignment for each
		\hline
		{} & {\textbf{Criteria:}}  \\
		\hline
		{Confirmed} & {(1) Appearing $\ge 4$ times in total failures.} \\
		{} & {(2)  Appearing at least once in the high-p failures.} \\
		\hline
		\hline
		{Suspected} & {(1) Appearing $\ge 2$ but $\le 4$ times in total failures.} \\
		{} & {(2) Does not satisfy the criteria for `confirmed'.} \\
		\hline
		\hline
		{Not misclassified} & {(1) Appearing $\le$ 1 time in the test of new models} \\
		\hline
		\hline
		
	\end{tabular}
	\caption{The criteria for selecting the suspected misclassified galaxies by the Galaxy Zoo project and purifying the training set.}
	\label{tab:confirm_criterion}
\end{table}
%figures
%figures

\begin{figure}
\begin{center}
\graphicspath{}
	%accuracy of each dataset and method.
	\includegraphics[width=0.8\columnwidth]{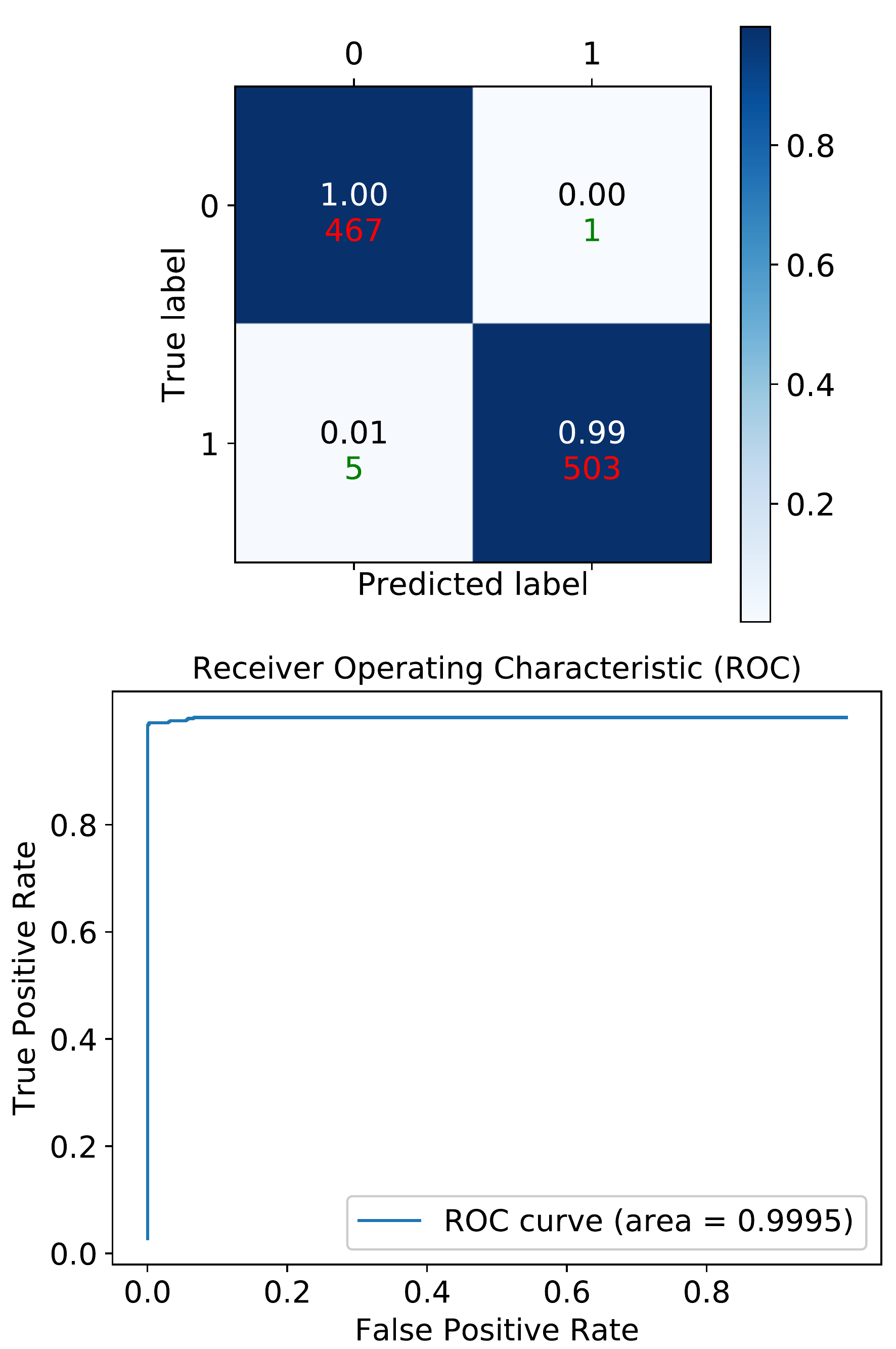}
  	\caption{The best testing result which we changed the label of the confirmed misclassified galaxies and excluded the suspected misclassified galaxies in both training and testing set. \textit{Top}: Confusion matrix. The `0' means Ellipticals and `1' represents Spirals. The colour bar shows the fraction of each true label (Galaxy Zoo), and the number shows the corresponding number of the fraction. \textit{Bottom}: The ROC curve of this testing result.}
    	\label{fig:corrected_best_result}
\end{center}
\end{figure}

The results are shown in Table~\ref{tab:final_result}. The first row of Table~\ref{tab:final_result} is the testing result excluding 8 suspected misclassified galaxies out of 1,000 testing galaxies. Compared this result with the results in Table~\ref{tab:criterion_clf}, our new models predict the highest accuracy, and end up having a resulting fewer number of uncertain type (about half the original number) than the previous results. Therefore, Fig.~\ref{fig:corrected_best_result} shows the best testing result in our study. In this result, we change the label of the confirmed misclassified galaxies and exclude the suspected misclassified galaxies in testing set. We obtain the accuracy of 0.994 for the best model within five reruns, and the average accuracy of five reruns is 0.991.

\begin{figure*}
\begin{center}
\graphicspath{}
	%accuracy of each dataset and method.
	\includegraphics[width=1.5\columnwidth]{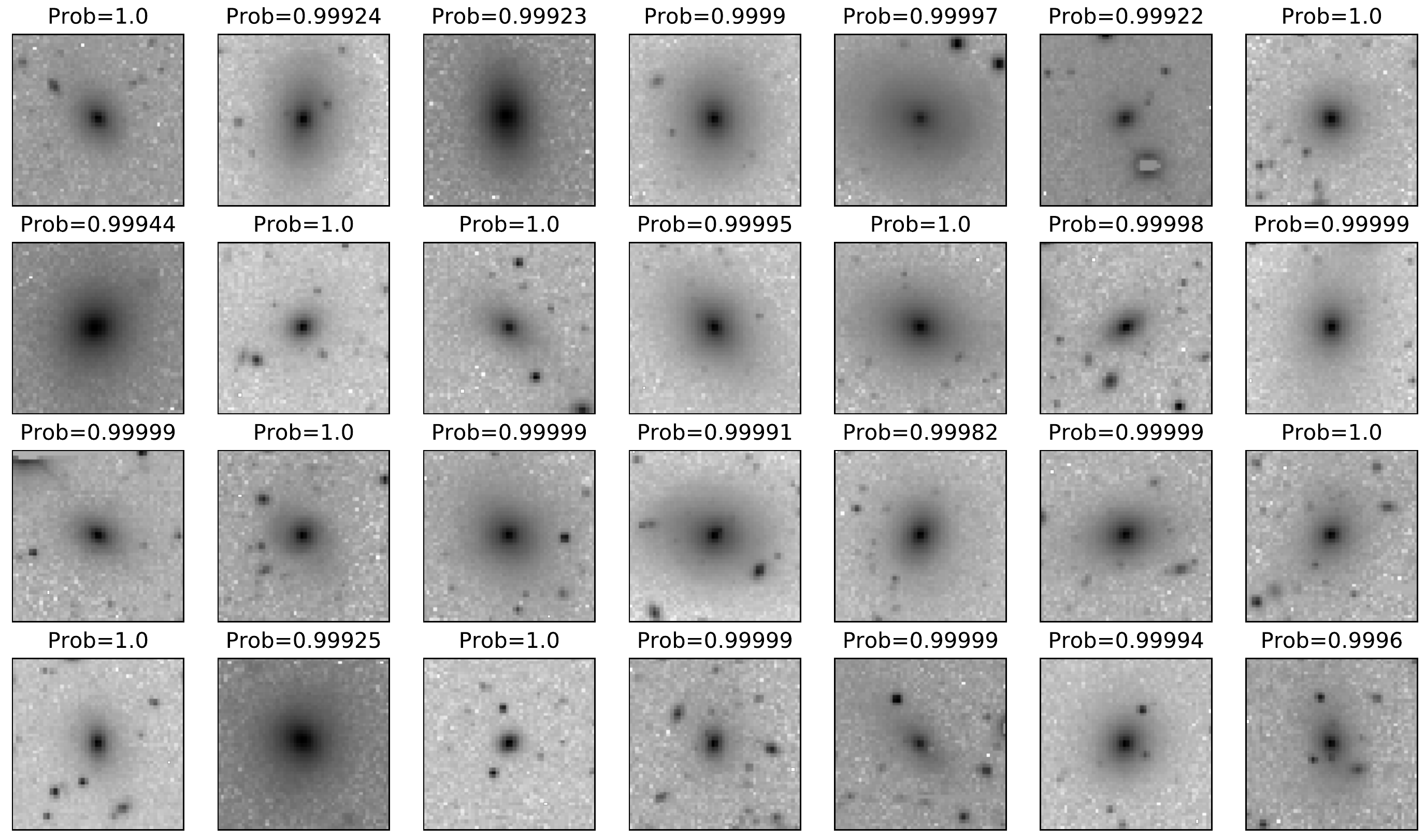}
  	\caption{Successful examples of classified Ellipticals. The `prob' on the top of the images show the predicted probability of being Ellipticals.}
    	\label{fig:success_E}
\end{center}
\end{figure*}

\begin{figure*}
\begin{center}
\graphicspath{}
	%accuracy of each dataset and method.
	\includegraphics[width=1.5\columnwidth]{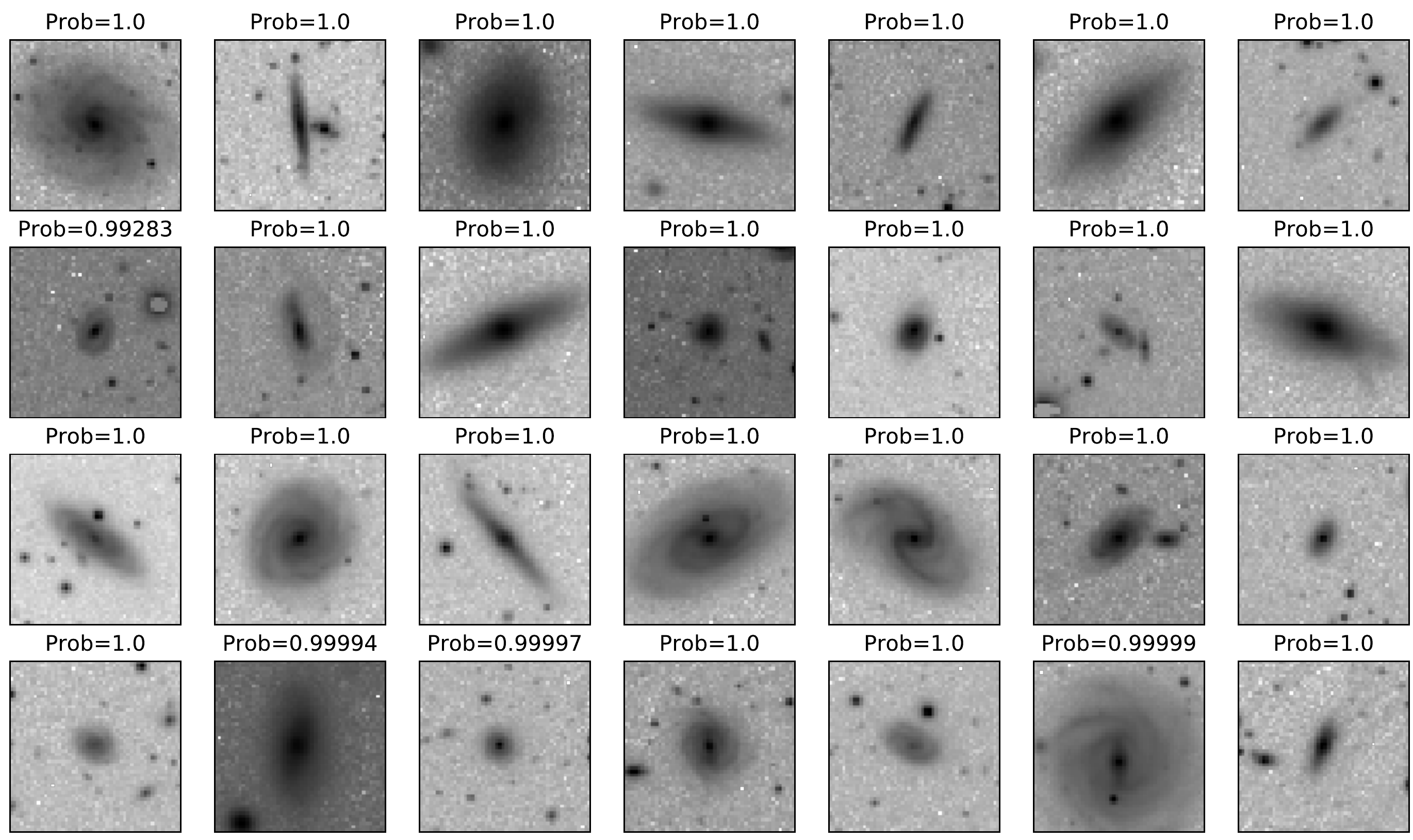}
  	\caption{Successful examples of the classified Spirals. The `prob' on the top of the images show the predicted probability of being Spirals.}
    	\label{fig:success_S}
\end{center}
\end{figure*}

The second and third rows of Table~\ref{tab:final_result} show the results including suspected galaxies which retain the initial label from the Galaxy Zoo in test and change the label of them to the opposite label, respectively. We have lower accuracy in these two conditions than the result of the first row. This indicates that part of our suspected galaxies have incorrect labels in Galaxy Zoo catalogue, and part of them are not, based on our CNN. Some examples of the successful classifications by the purified CNN training are shown in Fig.~\ref{fig:success_E} and Fig.~\ref{fig:success_S}.
\begin{table}
	%the final results after two times retrain
	\centering
	%\hspace*{-0.5cm}
	\begin{tabular}{cccccc} % four columns, alignment for each
		\hline
		{} & {accuracy} & {$R_{ 01 }$} & {${ N }_{ \text{classifiable} }$} & ${ N }_{ \text{uncetain} }$\\
		\hline
		\hline
		{No suspects} & {0.991$\pm$0.003} & {0.990} & {976} & {16} \\
		{with suspects} & {0.989$\pm$0.001} & {0.990} & {981} & {19} \\
		{label changed} & {0.987$\pm$0.003} & {0.986} & {981} & {19} \\
		\hline
		\hline
	\end{tabular}
	\caption{The testing result after using the purified training set. The meaning of each column are same as Table~\ref{tab:criterion_clf}. There are 8 suspected misclassified galaxies out of 1,000 testing galaxies. The first row is the testing result excluding suspected galaxies. The second row shows the result with the suspected galaxies which retain their initial labels from the Galaxy Zoo catalogue. The third row is the result with the suspected galaxies but their initial labels changed -- for instance, the label changes to Elliptical if the initial label was Spiral.}
	\label{tab:final_result}
\end{table}
%figures
%figures
%
%figures
%
%SECTION%
\section{Conclusions}
\label{sec:conclusions}
In this study, we have examined ten supervised machine learning methods to determine the most successful method for classifying galaxies into ellipticals and spirals using only pixel input on a single band ($i$-band). As part of the investigation, we have also tested how using rotated images with various angles of rotation with 10 degrees increments to augment our data influences on our classification. In addition, we also confirmed that the balance between the number ratio of each type is rather important when using pixel input in machine learning.

We show that the machine learning algorithms, Logistic Regression (LR) and Support Vector Machine (SVM) improve the performance of machine learning when combining with neural networks features, such as Restricted Boltzmann Machine (RBM). However, we find that using the image input along with the the Histogram of Oriented Gradient (HOG image) helps the performance in most methods, except for k-Nearest Neighbour (KNN). We also observe that the application of HOG images gives less help when combining with a neural network (e.g. LR+RBM, SVM+RBM, RF+RBM) because the RBM interlinks the HOG image features which have less information than the raw images. However, increasing the number of hidden layers and neurons qualitatively helps the connection between the HOG image features according to the performance of Multi-Layer Perceptron Classifier (MLPC) and Convolutional Neural Networks (CNN).

According to the Receiver Operating Characteristic (ROC) curve, the computing accuracy and the efficiency of each method, the performance of RF is comparable with a neural network (i.e. MLPC) with a faster computation time. In addition to RF, both the KNN and MLPC are alternative options can be considered when using pixel input because both of them have a relatively good accuracy with much less computing time than other conventional machine learning algorithms (e.g. LR, SVM) shown in this study (Table~\ref{tab:efficiency}). The most successful method within the ten methods we test is the Convolutional Neural Networks (CNN) with the combination input of raw images and HOG images and when using a balanced training data.  Through this we are able to reach an accuracy of $\sim$0.95 using $\sim$12,000 galaxies (including rotated images) as the initial training set. When using a classification criterion for the probability of the predicted type, $p>0.8$, we increase the accuracy to $\sim$0.97 and we are able to separate the classification into three types - Ellipticals, Spirals, and Uncertain. In the final test, when we apply the available maximum number of training data to train our CNN, and classified our testing galaxies by the criterion $p>0.8$, we reach a very high accuracy of $\sim$0.987 in the automated morphological classification of Ellipticals and Spirals.

In the discussion, we investigate the probable reasons for the failures in a small number of our classifications. We separate the failure into two situations - galaxies with high probabilities but still misclassified according to Galaxy Zoo, and galaxies with lower probabilities of being either elliptical or spiral. Most of galaxies in these two situations are repeated between the five reruns we do; therefore, these galaxies have some features in common which cause the difficulties within our CNN algorithm. 

We conclude that these `failures' are not true failures of the CNN. First of all, there is not a class for lenticular galaxy classification in the Galaxy Zoo catalogue, therefore, the confusion of lenticular galaxies with various labels cause difficulties to our CNN, resulting in low probability classifications for both ellipticals and spirals. Secondly, the better resolution (${ 0. }^{ \prime \prime  }263$ per pixel) and deeper depth ($i=$22.51) of DES data compared to the SDSS data reveals a more detailed structure of our sample of galaxies. Ultimately, this reveals incorrect labels from the Galaxy Zoo catalogue, due to the lower resolution and shallower depth of that data.   As a result we find a few misclassifications by the Galaxy Zoo project, identified through our machine learning. We find that about $2.5\%$ of the Ellipticals and Spirals are mislabelled out of $\sim 2,800$ galaxies from Galaxy Zoo. After correcting the labels of these confirmed misclassified galaxies by Galaxy Zoo, we reach an average accuracy of over 0.99 (0.994 in the best result within five reruns, Fig.~\ref{fig:corrected_best_result}) on the classification of Ellipticals and Spirals by our CNN.

In summary, the purpose of this paper is to pick the most successful machine learning method through pixel input for future usage in DES. With this method, we can quickly classify over millions of galaxies in DES data using a pre-trained model. Meanwhile, with current classification catalogues from other surveys and our own visual classification for galaxies in fainter bands, we can cross-validate and statistically analyse our classification by this optimal method on DES data. The most optimal method found amongst the 10 methods used in this paper is CNN. Ultimately, we will apply our CNN models trained by corrected labels of galaxies on DES data to build the largest morphological catalog ever with machine learning classifications. There is not a catalogue of morphological classification of galaxies for DES yet. Therefore, this catalogue as a reference will be useful for a comparison or further investigation with other studies. The binary classification in our paper has an advantage for direct blind tests of machine learning comparisons but otherwise has very limited application, therefore, we will also extend our algorithm to do more complicated morphological classifications of galaxies afterwards. 

In the longer term, we are developing the usage of Unsupervised Machine Learning (UML) for galaxy classification using pixel input. UML has no need for (much) pre-labelled data, so it can reduce the bias from human influences and interference as much as possible. At the same time it saves time which would otherwise be used to labelling data. With the development of UML and the Big Data from DES data, it will be very interesting to investigate the scenario of the evolution of galaxies and different possible classifications through machine learning. 
%##################################%
\section*{Acknowledgements}
Funding for the DES Projects has been provided by the U.S. Department of Energy, the U.S. National Science Foundation, the Ministry of Science and Education of Spain, the Science and Technology Facilities Council of the United Kingdom, the Higher Education Funding Council for England, the National Center for Supercomputing Applications at the University of Illinois at Urbana-Champaign, the Kavli Institute of Cosmological Physics at the University of Chicago, the Center for Cosmology and Astro-Particle Physics at the Ohio State University, the Mitchell Institute for Fundamental Physics and Astronomy at Texas A$\&$M University, Financiadora de Estudos e Projetos, Fundação Carlos Chagas Filho de Amparo à Pesquisa do Estado do Rio de Janeiro, Conselho Nacional de Desenvolvimento Científico e Tecnológico and the Ministério da Ciência, Tecnologia e Inovação, the Deutsche Forschungsgemeinschaft, and the Collaborating Institutions in the Dark Energy Survey.

The Collaborating Institutions are Argonne National Laboratory, the University of California at Santa Cruz, the University of Cambridge, Centro de Investigaciones Energéticas, Medioambientales y Tecnológicas-Madrid, the University of Chicago, University College London, the DES-Brazil Consortium, the University of Edinburgh, the Eidgenössische Technische Hochschule (ETH) Zürich, Fermi National Accelerator Laboratory, the University of Illinois at Urbana-Champaign, the Institut de Ciències de l'Espai (IEEC/CSIC), the Institut de Física d'Altes Energies, Lawrence Berkeley National Laboratory, the Ludwig-Maximilians Universität München and the associated Excellence Cluster Universe, the University of Michigan, the National Optical Astronomy Observatory, the University of Nottingham, The Ohio State University, the University of Pennsylvania, the University of Portsmouth, SLAC National Accelerator Laboratory, Stanford University, the University of Sussex, Texas A$\&$M University, and the OzDES Membership Consortium.

Based in part on observations at Cerro Tololo Inter-American Observatory, National Optical Astronomy Observatory, which is operated by the Association of Universities for Research in Astronomy (AURA) under a cooperative agreement with the National Science Foundation.

The DES data management system is supported by the National Science Foundation under grant numbers AST-1138766 and AST-1536171. The DES participants from Spanish institutions are partially supported by MINECO under grants AYA2015-71825, ESP2015-66861, FPA2015-68048, SEV-2016-0588, SEV-2016-0597, and MDM-2015-0509, some of which include ERDF funds from the European Union. IFAE is partially funded by the CERCA programme of the Generalitat de Catalunya. Research leading to these results has received funding from the European Research Council under the European Union's Seventh Framework Program (FP7/2007-2013) including ERC grant agreements 240672, 291329, and 306478. We acknowledge support from the Australian Research Council Centre of Excellence for All-sky Astrophysics (CAASTRO), through project number CE110001020, and the Brazilian Instituto Nacional de Ciencia e Tecnologia (INCT) e-Universe (CNPq grant 465376/2014-2).

%%%%%%%%%%%%%%%%%%%%%%%%%%%%%%%%%%%%%%%%%%%%%%%%%%



\bibliographystyle{mnras}
\bibliography{ms}

%%%%%%%%%%%%%%%%%%%%%%%%%%%%%%%%%%%%%%%%%%%%%%%%%%

%%%%%%%%%%%%%%%%% APPENDICES %%%%%%%%%%%%%%%%%%%%%
\appendix
\section{Support Vector Machine}
\label{sec:SVM_appendix}
\begin{figure}
\begin{center}
\graphicspath{}
	%SVM 2D illustration
	\includegraphics[width=\columnwidth]{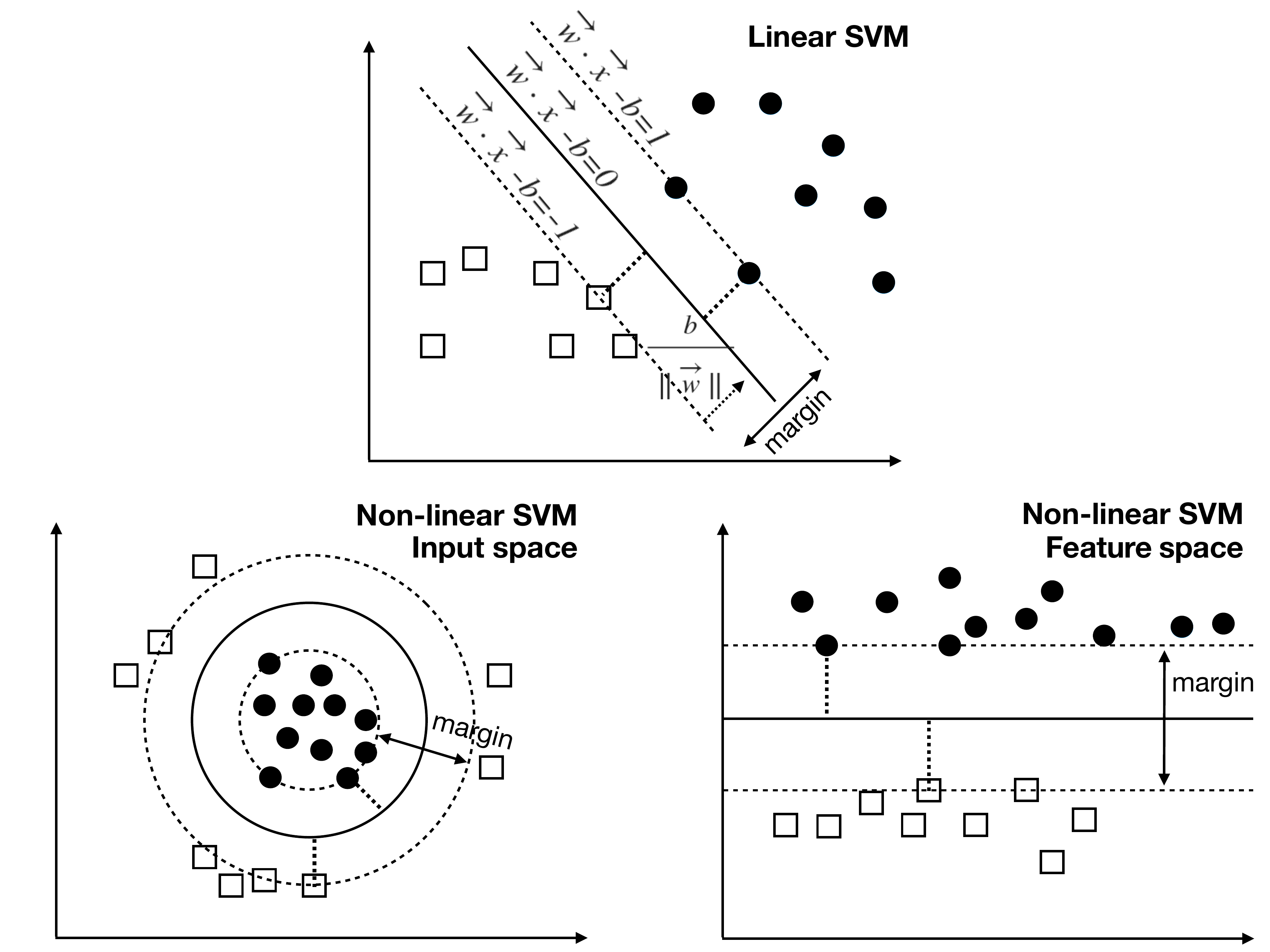}
   	\caption{Illustration of the linear and non-linear SVM method. Different markers represent two different classifications. {\it{Top}}: linear SVM. {\it{Bottom Left}}: non-linear SVM in input space. {\it{Bottom Right}}: non-linear SVM in feature space (kernel space).}
    	\label{fig:svm_illustration}
\end{center}
\end{figure}

Support Vector Machine (SVM) algorithm is to find a hyperplane defined as below, 
%equation
\begin{equation}
    	\vec { w } \cdot \vec { x } -b=0,
\end{equation}
\noindent where $\vec {w}$ is  a weighted vector, $\vec {x}$ is the input data, and $b$ is the bias, with the maximum distance to the nearest data for each type (\textit{support vector}): $\left| \vec { w } \cdot \vec { x } -b \right| =1$ \citep{Vapnik1995, Cortes1995}. For example (See the top of Fig.~\ref{fig:svm_illustration}), in 2-class classification, $\left\{ {\vec{{ x }_{ j }}},{ y }_{ j } \right\}  $,  ${\vec{{ x }_{ j }}}$ is a vector which represents input data, and ${y}_{j}$ represents the classification. The $j$ means the $j$-th data. ${ y }_{ j }\in \left\{ 1(\text{circle}),-1(\text{square}) \right\}$. While the parameter $\frac { b }{ \left\| \vec { w }  \right\|  } $ determines the distance between the hyperplane to the support vectors, finding the maximum of this parameter is finding the minimum $\left\| \vec { w }  \right\| $. After determining the decision boundary, data above the boundary: $\vec { w } \cdot \vec { x } -b\ge 1$ is classified as a circle, the below one: $\vec { w } \cdot \vec { x } -b\le -1$ is classified as a square.

When using a non-linear SVM, the algorithm uses a kernel function $K$ to the data: $\left( { \vec{x} },\vec{{ x }}^{ ' } \right) \rightarrow K\left( { \vec{x} },\vec{{ x }}^{ ' } \right) $ to map the data. The bottom of Fig.~\ref{fig:svm_illustration} shows a 2D illustration of an example of non-linear SVM with a circular transformation. In this example, we assume each point is $\left( { a }_{ k },{ b }_{ k } \right) $, and we transform the data into a new feature space which is defined as $c=\sqrt { { a }_{ k }^{ 2 }+{ b }_{ k }^{ 2 } } $ (circular transformation); therefore, the decision boundary is shown as the circular shape in the input space (i.e. $a-b$ space), but shown lines in feature space ($c$ space). 

There are two standard regularisation parameters for SVM: C-SVM and Nu-SVM \citep{Scholkopf2001} methods. Both C and Nu are the parameter of regularisation which are related to the number of support vectors and the number of misclassification. The range of C can be any positive value, but the range of Nu is limited to 0 and 1 which is easier to control.
%figure


%%%%%%%%%%%%%%%%%%%%%%%%%%%%%%%%%%%%%%%%%%%%%%%%%%

% Don't change these lines
\bsp	% typesetting comment
\label{lastpage}
\end{document}